\documentclass[aip, jcp, twocolumn, floatfix, superscriptaddress, reprint, 10pt]{revtex4-1}
\usepackage{physics}
\usepackage{xparse}

\newcommand{\mbf}[1]{\ensuremath{\mathbf{#1}}}

\NewDocumentCommand{\rep}{s d<| d|>}{%
\IfBooleanTF{#1}{
   \IfValueTF{#2}{
       \IfValueTF{#3}{\braket{#2}{#3}}{\bra{#2}}
       }{
       \IfValueTF{#3}{\ket{#3}}{}
       }
   }{
   \IfValueTF{#2}{
       \IfValueTF{#3}{\braket*{#2}{#3}}{\bra*{#2}}
       }{
       \IfValueTF{#3}{\ket*{#3}}{}
       }
   }
}

\NewDocumentCommand{\rbra}{sm}{\IfBooleanTF{#1}{\rep*<#2|}{\rep<#2|}}
\NewDocumentCommand{\rket}{sm}{\IfBooleanTF{#1}{\rep*|#2>}{\rep|#2>}}
\NewDocumentCommand{\rbraket}{smm}{\IfBooleanTF{#1}{\rep*<#2||#3>}{\rep<#2||#3>}}

\NewDocumentCommand{\field}{o m e{_} e{^} o e{_} e{^}}{
\IfValueTF{#5}{\overline{
  #2\IfValueT{#3}{_#3}\IfValueT{#4}{^{\otimes #4}} %
  \otimes
  #5\IfValueT{#6}{_#6}\IfValueT{#7}{^{\otimes #7}} %
  \IfValueT{#1}{;#1}
}}{
  \IfValueTF{#4}{\overline{
     #2\IfValueT{#3}{_#3}\IfValueT{#4}{^{\otimes #4}}
     \IfValueT{#1}{;#1}
  }}
  {#2\IfValueT{#3}{_#3}}
}
}

\NewDocumentCommand{\frho}{o e{_} e{^}}{
\field[#1]{\rho}_{#2}^{#3}
}

\newcommand{\e}{a}  %

\newcommand{\br}{\mbf{r}}
\newcommand{\bx}{\mbf{x}}
\newcommand{\bxhat}{\hat{\mbf{x}}}

\NewDocumentCommand{\ex}{e_}{
\IfValueTF{#1}{\e_{#1}\bx_{#1}}{\e\bx}
}  %

\NewDocumentCommand{\lm}{e_}{
\IfValueTF{#1}{l_{#1}m_{#1}}{lm}
}
\NewDocumentCommand{\nlm}{e_}{
\IfValueTF{#1}{n_{#1}\lm_{#1}}{n\lm}
}
\NewDocumentCommand{\enlm}{e_}{
\IfValueTF{#1}{\e_{#1}\nlm_{#1}}{\e\nlm}
}
\NewDocumentCommand{\en}{e_}{
\IfValueTF{#1}{\e_{#1}n_{#1}}{\e n}
}
\NewDocumentCommand{\nlk}{e_}{
\IfValueTF{#1}{n_{#1}l_{#1}k_{#1}}{nlk}
}
\NewDocumentCommand{\enlk}{e_}{
\IfValueTF{#1}{\e_{#1}\nlk_{#1}}{\e\nlk}
}
\NewDocumentCommand{\enl}{e_}{
\IfValueTF{#1}{\en_{#1}l_#1}{\en l}
}
\NewDocumentCommand{\nl}{e_}{
\IfValueTF{#1}{n_{#1}l_#1}{n l}
}

\NewDocumentCommand{\nnl}{s}{
\IfBooleanTF{#1}{n_1 n_2 l}{n_1; n_2; l}
}
\NewDocumentCommand{\ennl}{s}{
\IfBooleanTF{#1}{\en_1 \en_2 l}{\en_1; \en_2; l}
}

\NewDocumentCommand{\gslm}{s}{
\IfBooleanTF{#1}{\sigma\lambda\mu}{\sigma;\lambda\mu}
}

\newcommand{\revadd}[1]{#1}
\newcommand{\revaddfig}[1]{#1}

\newcommand{\sigmag}{\sigma_\text{G}}
\newcommand{\lmax}{{l_\text{max}}}
\newcommand{\nmax}{{n_\text{max}}}

\newcommand{\nf}{m}
\newcommand{\ns}{n}

\newcommand{\bX}{\mathbf{X}}
\newcommand{\bXt}{{\tilde{\bX}}}

\newcommand{\bphi}{\boldsymbol{\phi}}
\newcommand{\bK}{\mathbf{K}}

\newcommand{\bP}[2]{\mathbf{P}_{{#1}{#2}}}
\newcommand{\bQ}[2]{\mathbf{Q}_{{#1}{#2}}}

\newcommand{\bU}{\mathbf{U}}
\newcommand{\bV}{\mathbf{V}}
\newcommand{\bSIG}{\boldsymbol{\Sigma}}
\newcommand{\bLAM}{\boldsymbol{\Lambda}}

\newcommand{\train}{\mathrm{train}}
\newcommand{\test}{\mathrm{test}}
\newcommand{\CD}{\mathcal{D}}
\newcommand{\CDtr}{\mathcal{D}_\train}
\newcommand{\CDte}{\mathcal{D}_\test}
\newcommand{\CF}{\mathcal{F}}
\newcommand{\CH}{\mathcal{H}}
\newcommand{\BR}{\mathbb{R}}

\newcommand{\GFRE}{\operatorname{GFRE}}
\newcommand{\GFRD}{\operatorname{GFRD}}
\newcommand{\LFRE}{\operatorname{LFRE}}

\usepackage{physics}

\newcommand{\D}[2][]{\ensuremath{\mathop{}\!\mathrm{d}^{#1}{#2}\,}}

\newcommand{\SOthree}{ {SO(3)} }
\newcommand{\zh}{z_{\ce{H}}}

\makeatletter
\let\oldket\ket
\def\ket{\@ifstar{\oldket}{\oldket*}}
\let\oldbra\bra
\def\bra{\@ifstar{\oldbra}{\oldbra*}}
\let\oldev\ev
\def\ev{\@ifstar{\oldev}{\oldev*}}
\makeatother
\usepackage{bm}
\usepackage{graphicx}
\usepackage[usenames]{color}
\usepackage{microtype}
\usepackage{amsmath}
\usepackage{amssymb}
\usepackage{xspace}
\usepackage{comment}
\usepackage{footnote}
\usepackage[version=4]{mhchem}
\usepackage{soul}
\usepackage{relsize}
\usepackage{hhline}
\usepackage{verbatim}
\usepackage[normalem]{ulem}
\usepackage[inline]{enumitem}
\usepackage{caption}

\newcommand{\argmin}[1]{\underset{#1}{\operatorname{arg}\,\operatorname{min}}\;}

\begin{document}

\title{The role of feature space in atomistic learning}

\author{Alexander Goscinski}
\affiliation{Laboratory of Computational Science and Modeling, IMX, \'Ecole Polytechnique F\'ed\'erale de Lausanne, 1015 Lausanne, Switzerland}

\author{Guillaume Fraux}
\affiliation{Laboratory of Computational Science and Modeling, IMX, \'Ecole Polytechnique F\'ed\'erale de Lausanne, 1015 Lausanne, Switzerland}

\author{Giulio Imbalzano}
\affiliation{Laboratory of Computational Science and Modeling, IMX, \'Ecole Polytechnique F\'ed\'erale de Lausanne, 1015 Lausanne, Switzerland}

\author{Michele Ceriotti}
\email{michele.ceriotti@epfl.ch}
\affiliation{Laboratory of Computational Science and Modeling, IMX, \'Ecole Polytechnique F\'ed\'erale de Lausanne, 1015 Lausanne, Switzerland}
\begin{abstract}
Efficient, physically-inspired descriptors of the structure and composition of molecules and materials play a key role in the application of machine-learning techniques to atomistic simulations.
The proliferation of approaches, as well as the fact that each choice of features can lead to very different behavior depending on how they are used, e.g. by introducing non-linear kernels and non-Euclidean metrics to manipulate them, makes it difficult to objectively compare different methods, and to address fundamental questions on how one feature space is related to another.
In this work we introduce a framework to compare different sets of descriptors, and different ways of transforming them by means of metrics and kernels, in terms of the structure of the feature space that they induce. We define diagnostic tools to determine whether alternative feature spaces contain equivalent amounts of information, and whether the common information is substantially distorted when going from one feature space to another.
We compare, in particular, representations that are built in terms of $n$-body correlations of the atom density, quantitatively assessing the information loss associated with the use of low-order features. We also investigate the impact of different choices of basis functions and hyperparameters of the widely used SOAP and Behler-Parrinello features, and investigate how the use of non-linear kernels, and of a Wasserstein-type metric, change the structure of the feature space in comparison to a simpler linear feature space.\end{abstract}

\maketitle

\section{Introduction}

The construction of efficient and insightful descriptors of atomic configurations has been one of the focal points of the development of data-driven applications for atomic-scale modeling~\cite{behl-parr07prl,bart+10prl,rupp+12prl,bart+13prb,de+16pccp,eick+17nips,huo2017unified,fabe+17jctc, chmiela2018towards,zhan+18prl,will+19jcp,drau19prb,chri+20jcp,van2020regularised, Ghiringhelli2015, zhu+16jcp, Gallet2013}.
Two of the core ideas that underlie most of the existing schemes are the use of an atom-centred description that is particularly well-suited to model additive, extensive properties, and the incorporation of geometric and atom permutation symmetries. While incorporation of symmetries makes representations much more data efficient, it raises subtle issues of whether the mapping from structure to descriptor is injective or not~\cite{bart+13prb,vonl+15ijqc, pozd+20prl}.
Many of the structural representations that fulfill these symmetry requirements are closely related to one another, corresponding to projections of $n$-body correlations of the atom density~\cite{will+18pccp,will+19jcp}. Yet, comparing them is not straightforward. When used to build an interatomic potential, or to predict another atomic-scale property, representations are used together with different supervised learning schemes, so it is difficult to disentangle the interplay of descriptor, regression method, and target property that combine to determine the accuracy and computational cost of the different methods.~\cite{zuo+20jpcl}
Juxtaposing alternative choices of representations is complicated by the fact that non-linear transformations are often applied as a part of the data processing algorithm, and so it would be equally important to be able to analyze the effect of these transformations.
Efforts to compare different choices of descriptors have been mostly focused this far on a comparison of compressibility~\cite{helf+20mlst,onat+20jcp}, their ability to represent atomic structures uniquely~\cite{moussa2012comment,bart+13prb,sadeghi2013metrics,pozd+20prl}, their role in constructing a metric~\cite{zhu+16jcp,de+16pccp} and their sensitivity to perturbations of the atomic structure~\cite{pars+20mlst, onat+20jcp}.

Here we propose a strategy to compare feature spaces both in terms of their mutual information content -- which we define transparently as the ability to linearly or non-linearly reconstruct each other -- and in terms of the amount of deformation that has to be applied to match the common information between the two.
Note that the definition we use here differs from that used in information-theoretical treatments, based on Shannon entropy -- which is difficult to compute in high dimensions~\cite{torkkola2003feature}, and does not reflect as naturally the behavior of different features when used in the context of atomistic machine learning.

We demonstrate this strategy by applying it to elucidate several issues related to the behavior of density-based representations.
First, we investigate the role of the basis and of the density smearing in the practical implementation of 3-body density correlation features; we then estimate the loss of information that one incurs by truncating the description to low body-order of correlations;  finally, we discuss the role of the metric used to compare two structures, by testing the commonly used Euclidean distance against kernel-induced and Wasserstein-type metrics. We provide an open-source implementation of functions to compute these quantities at \url{https://github.com/cosmo-epfl/scikit-cosmo}.

\section{Comparing feature spaces}

Consider a dataset $\CD=\left\{x_i\right\}$ containing $\ns$ items. For a given choice of features $\CF$, each item is described by an $\nf_\CF$-dimensional feature vector $\bx_i$. As a whole, the dataset is described by a feature matrix $\bX_\CF^\CD \in \BR^{\ns\times\nf_\CF}$.
We consider all of the feature matrices in this work to be standardized, i.e. centred and scaled so as to have zero mean and unit variance for the selected data set.
Consider a second featurization $\CF'$. We want to be able to compare the behavior of different choices of feature spaces when representing the dataset $\CD$, e.g. which of two sets of features have more expressive power, and how much distorted is one representation relative to the other.

\begin{figure*}[tbhp]
\centering
\includegraphics[width=1.0\linewidth]{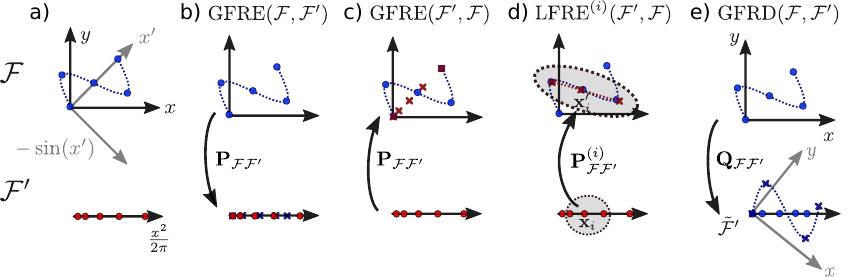}
\caption{\revadd{A schematic representation of the different measures of feature-space dissimilarity we introduce in this work, discussed from left to right.
The figure considers a dataset containing five samples, embedded in a two-dimensional feature space $\CF$ and a one-dimensional feature space $\CF'$.
As shown, the relationship between the two embeddings can involve arbitrary linear and non-linear transformations (panel a).
The global feature space reconstruction error (GFRE) defined in Eq.~\eqref{eq:GFRE} amounts to finding the best linear mapping between the two feature spaces.
This measure is not symmetric: in this example the $\GFRE(\CF,\CF')$ (panel b) is smaller than the $\GFRE(\CF',\CF)$ (panel c), since $\CF$ contains an additional nonzero dimension.
The local version of the reconstruction error (LFRE) defined in Eq.~\eqref{eq:LFRE} makes it possible to probe whether a non-linear map exists between the two spaces: in this case, the sinus function can be approximated in  neighbourhood of each sample $\bx_i$ by a linear map $\bP{\CF}{\CF'}^{(i)}$  defined in Eq.~\eqref{eq:xtilde_lle}; by treating each neighbourhood separately  it is possible to achieve a low $\LFRE(\CF',\CF)$ (panel d).
Finally, the global feature space reconstruction distortion (GFRD) defined in Eq.~\eqref{eq:GFRD} determines whether the two featurizations are connected by an orthogonal transformation $\bQ{\CF}{\CF'}$ defined in Eq.~\eqref{eq:proc-ff1}, by finding the best alignment between $\CF$ and the approximation of $\CF'$ that can be obtained as a linear projection of $\CF$.
Even though $\GFRE(\CF,\CF')$ is small, one of the components of $\CF$ is scaled down to zero, resulting in a large value of $\GFRD(\CF,\CF')$ (panel e).}}
\label{fig:gfr-scheme}
\end{figure*}

\subsection{Global feature space reconstruction error}

As a simple, easily-interpretable measure of the relative expressive power of $\CF$ and $\CF'$, we introduce the global feature space reconstruction error $\GFRE^\CD(\CF,\CF')$, defined as the mean-square error that one incurs when using the feature matrix $\bX_\CF$ to linearly regress $\bX_{\CF'}$.
In this work we compute the $\GFRE$ by a 2-fold split of the dataset, i.e. compute the regression weights $\bP{\CF}{\CF'}$ over a train set $\CDtr$ composed of half the entries in $\CD$,
\begin{equation}
\begin{split}
\bP{\CF}{\CF'}= &
\argmin{\bP{}{}\in\BR^{\nf_{\CF}\times\nf_{\CF'}}}
\norm{\bX^{\CDtr}_{\CF'} - \bX^{\CDtr}_{\CF} \bP{}{}  }\\
=&\left({\bX_{\CF}^{\CDtr}}^T \bX_{\CF}^{\CDtr}\right)^{-1}
(\bX_{\CF}^{\CDtr})^T\bX_{\CF'}^{\CDtr}
\end{split}\label{eq:proj-ff1}
\end{equation}
and then compute the error over the remaining test set $\CDte$
\begin{equation}
\GFRE^\CD(\CF,\CF') = \sqrt{{\norm{\bX^{\CDte}_{\CF'} - \bX^{\CDte}_{\CF} \bP{\CF}{\CF'}  }^2}/\ns_\test},
\label{eq:GFRE}
\end{equation}
averaging, if needed, over multiple random splits.  The $\GFRE$ is a positive quantity, which is equal to zero when there is no error in the reconstruction, and that is usually bound by one\footnote{This is due to the fact that feature matrices are standardized, and so $\norm{\bX^{\CDte}_{\CF'}}/\ns_\test $ is of the order of one}.
For numbers of features larger than $\ns_\train$, the covariance matrix is not full rank, and one needs to compute a pseudoinverse. Without loss of generality, one can regularize the regression to stabilize the calculation. In this paper, we computed the pseudoinverse by means of a singular value decomposition, and we determined the optimal regularization in terms of the truncation of the singular value spectrum, using 2-fold cross-validation over the training set to determine the optimal truncation threshold. Often, it is also useful to observe the behavior of the $\GFRE$ in the absence of any regularization: overfitting is in itself a signal of the instability of the mapping between feature spaces.
In general, $\GFRE^\CD(\CF,\CF')$ is not symmetric. If $\GFRE^\CD(\CF,\CF')\approx\GFRE^\CD(\CF',\CF)\approx 0$, $\CF$ and $\CF'$ contain similar types of information; if $\GFRE^\CD(\CF,\CF')\approx 0$, while $\GFRE^\CD(\CF',\CF)>0$, one can say that $\CF$ is more descriptive than $\CF'$: this is the case, for instance, one would observe if $\CF'$ consists of a sparse version of $\CF$, with some important and linearly-independent features removed; finally, if  $\GFRE^\CD(\CF,\CF')\approx\GFRE^\CD(\CF',\CF)>0$, the two feature spaces contain different, and complementary, kinds of information and it may be beneficial to combine them to achieve a more thorough description of the problem.

\subsection{Global feature space reconstruction distortion}

The feature space reconstruction error gives insights into whether a feature space can be inferred by knowledge of a second one. However, having both a small $\GFRE^\CD(\CF,\CF')$ and $\GFRE^\CD(\CF',\CF)$ does not imply two feature spaces are identical. Even though they contain similar amounts of information, one feature space could give more emphasis to some features compared to the other, which can eventually result in different performance when building a model.
To assess the amount of distortion of $\CF'$ relative to $\CF$, we introduce the global feature space reconstruction distortion $\GFRD^\CD(\CF,\CF')$.
To evaluate it, we first compute the singular value decomposition of the projector Eq.~\eqref{eq:proj-ff1}, $\bP{\CF}{\CF'}\approx\bU \bSIG \bV^T$, and then use it to reduce the two feature spaces to a common basis, in which the reconstruction error is zero, because the residual has been discarded
\begin{equation}
\bXt_{\CF} = \bX_{\CF}\bU \quad  \bXt_{\CF'} = \bXt_{\CF} \bSIG.
\end{equation}
When the second feature space $\CF'$ has a lower dimensionality than $\CF$, some combinations of the starting features are not used to compute $\tilde{\CF}'$.
In this case, we pad $\bSIG$ with zeros, so that $\tilde{\CF}'$ has the same dimensionality $\nf_{\CF}$ as the starting space. This choice ensures that the GFRD takes the same value it would have in the case $\CF'$ had the same dimensionality as $\CF$, but lower rank.
In the opposite case, with $\nf_{\CF}<\nf_{\CF}'$, padding $\bSIG$ and $\bU$ with zeros, or truncating $\bV$, yields the same $\GFRD$.

We can then address the question of whether $\bXt_{\CF}$ and $\bXt_{\CF'}$ are linked by a unitary transformation (in which case the $\GFRD$ should be zero), or there is a distortion involved.
A possible answer involves solving the orthogonal Procrustes problem~\cite{scho66pm} -- i.e. finding the orthogonal transformation that ``aligns'' as well as possible $\bXt_{\CF}$ to $\bXt_{\CF'}$:
\begin{equation}
\begin{split}
\bQ{\CF}{\CF'} =& \argmin{\bQ{}{} \in \mathbb{U}^{\nf\times\nf}}
\norm{\bXt^{\CDtr}_{\CF'} - \bXt^{\CDtr}_{\CF} \bQ{}{}  }\\
=&\tilde{\bU}\tilde{\bV}^T,
\end{split}\label{eq:proc-ff1}
\end{equation}
where $\tilde{\bU}\tilde{\bSIG}\tilde{\bV}^T = (\bXt_{\CF}^{\CDtr})^T \bXt^{\CDtr}_{\CF'}$ .
The amount of distortion can then be computed by assessing the residual on the test set,
\begin{equation}
\GFRD^\CD(\CF,\CF') = \sqrt{{\norm{{\bXt}^{\CDte}_{\CF'} - {\bXt}^{\CDte}_{\CF} \bQ{\CF}{\CF'}  }^2}/\ns_\test}. \label{eq:GFRD}
\end{equation}
If desired, the error can be averaged over multiple random splits of the reference data set $\CD$.

\subsection{Local feature space reconstruction error}

A downside of the global feature comparison schemes introduced above is that the linear nature of the regression means that they cannot detect if $\CF$ and $\CF'$ contain analogous information, but differ by a non-linear transformation.
In the next Section we discuss how one can generalize the schemes to use kernel features, that can also be used to detect non-linear relationships between the original feature spaces.
\newcommand{\CDki}{\CD_{k-\text{neigh}}^{(i)}}
An alternative approach is to compute a local version of the feature space reconstruction error, $\LFRE^\CD(\CF,\CF')$, loosely inspired by locally-linear embedding~\cite{rowe-saul00science}.
To compute the LFRE, a local regression is set up, computed in the $k$-neighbourhood $\CDki$ around sample $i$  -- the set of $k$ nearest neighbours of sample $i$, based on the Euclidean distance between the samples in $\CF$ -- to reproduce the $\CF'$ features using $\CF$ as input features, centred around their mean values $\bar{\bx}_{\CF'}$ and $\bar{\bx}_{\CF}$.

A local embedding of $\bx_i$ is determined as
\begin{equation}
\tilde{\bx}'_i = \bar{\bx}_{\CF'} + (\bx_i - \bar{\bx}_{\CF})\bP{\CF}{\CF'}^{(i)},\label{eq:xtilde_lle}
\end{equation}
where $\bP{\CF}{\CF'}^{(i)}$ contains the regression weights computed from $\CDki$.
The local feature space reconstruction error is given by the residual discrepancy between the $\CF'$ counterpart of the $i$-th point and its local embedding~\eqref{eq:xtilde_lle}:
\begin{equation}
\LFRE^\CD(\CF,\CF') = \sqrt{\sum_i \norm{\bx'_i - \tilde{\bx}'_i}^2/{\ns_\test}}.
\label{eq:LFRE}
\end{equation}
Inspecting the error associated with the reconstruction of individual points can reveal regions of feature space for which the mapping between $\CF$  and $\CF'$ is particularly problematic.
Similarly, one can compute a local version of $\GFRD$, that could be useful to detect strong local distortions that might indicate the presence of a singularity in the mapping between two feature spaces.

\subsection{Bending space: comparing induced feature spaces}

It is often possible to substantially improve the performance of regression or dimensionality reduction algorithms, without explicitly changing the feature vectors. This can be achieved by introducing a (non-linear) similarity measure to compare $\bx_i$, which takes the form of a kernel function $k(\bx,\bx')$, or a dissimilarity measure which takes the form of a distance $d(\bx,\bx')$.

Let us recall that a positive-definite kernel induces a kernel distance by the relation\cite{scholkopf2001kernel}
\begin{equation}
    d_k(\bx,\bx')^2 = k(\bx,\bx) + k(\bx',\bx') -2k(\bx,\bx'),
    \label{eq:kernel_distance_relation}
\end{equation}
and that any negative-definite distance can be used to build positive-definite kernels such as the substitution kernel\cite{haasdonk2004learning}
\begin{equation}
k_d^{\mathbf{x}_0}(\bx,\bx') = -\frac12(d(\bx,\bx')^2 - d(\bx,\bx_0)^2 - d(\bx_0,\bx')^2),  \bx_0\in\CF \label{eq:k-0}
\end{equation}
or the radial basis function (RBF) kernel
\begin{equation}
k_d^{\textrm{RBF}}(\bx,\bx') = \exp(-\gamma d(\bx,\bx')^2),\quad \gamma\in\BR_+ \label{eq:k-rbf}
\end{equation}

A positive definite kernel induces a feature space $\CH$, commonly known as reproducing kernel Hilbert space (RKHS), in which the similarity measure can be expressed as a dot product:
\begin{equation}
    k(\bx,\bx') = \langle\bphi(\bx)|\bphi(\bx')\rangle,\quad \bx,\bx'\in\mathcal{F},\,\, \bphi:\mathcal{F}\rightarrow\CH.
\end{equation}
While in general $\bphi(\bx)$ is not known, for a given dataset $\CD$ it is possible to approximate the RKHS features by using a kernel principal component analysis~\cite{scholkopf1997kernel}. Since linear regression in RKHS features is equivalent to kernel ridge regression, we simply use kernel features computed on the training  dataset $\CDtr$ to reduce the problem of comparing kernel (or distance) induced features to that of comparing explicit features, and use $\GFRE$ and $\GFRD$ as defined in Eqs.~\eqref{eq:GFRE} and~\eqref{eq:GFRD}.
It is possible to re-formulate these measures in an explicit kernelized form, as well as to compute low-rank approximations of the kernel to reduce the computational cost for very large datasets (see e.g. Ref.~\citenum{helf+20mlst} for a pedagogic discussion). In this paper we simply use the explicit RKHS features, that can be obtained by diagonalizing the kernel matrix $\bK = \bU \bLAM \bU^T$, with $K_{ij} = k(\bx_i, \bx_j)$, and defining
\begin{equation}
\bX_\CH = \bU \bLAM^{-1/2},
\end{equation}
which is then standardized as we do for any other set of features.
To define a feature space associated with a metric, rather than a kernel, we first center the squared distance matrix (which is equivalent to computing a substitution kernel analogous to Eq.~\eqref{eq:k-0}) and then proceed similarly by diagonalizing the resulting matrix.

\subsection{Dataset selection}

We use four different datasets, chosen to emphasize different aspects of the problem of representing atomic structures: A \textit{random methane} dataset consisting of different random displacements of the four hydrogen atoms around the central carbon atom to cover the complete configurational space of \ce{CH4} structures; A \textit{carbon} dataset of approximately 10'000 minimum energy carbon structures, obtained as the result of ab initio random structure search~\cite{pick-need11jpcm, pickard-carbon}, as an example for a realistic dataset of condensed phase structures; A \textit{degenerate methane} dataset composed of two groups of methane structures (which we refer to as $\mathcal{X}^+$ and $\mathcal{X}^-$), each associated with a 2D manifold parameterised by two parameters $(u,v)$: structures with $v=0$ in the two manifolds have exactly the same C-centred 3-body correlations, despite being different (as discussed in Ref.~\citenum{pozd+20prl});  A \textit{displaced methane} dataset, which consists in an ideal, tetrahedral \ce{CH4} geometry with one hydrogen atom pulled away from the central carbon atom, as an example of a set of structures that are distinguished by a clearly identifiable structural feature, here the \ce{C-H} distance.

\begin{figure*}
    \includegraphics[width=0.9\linewidth]{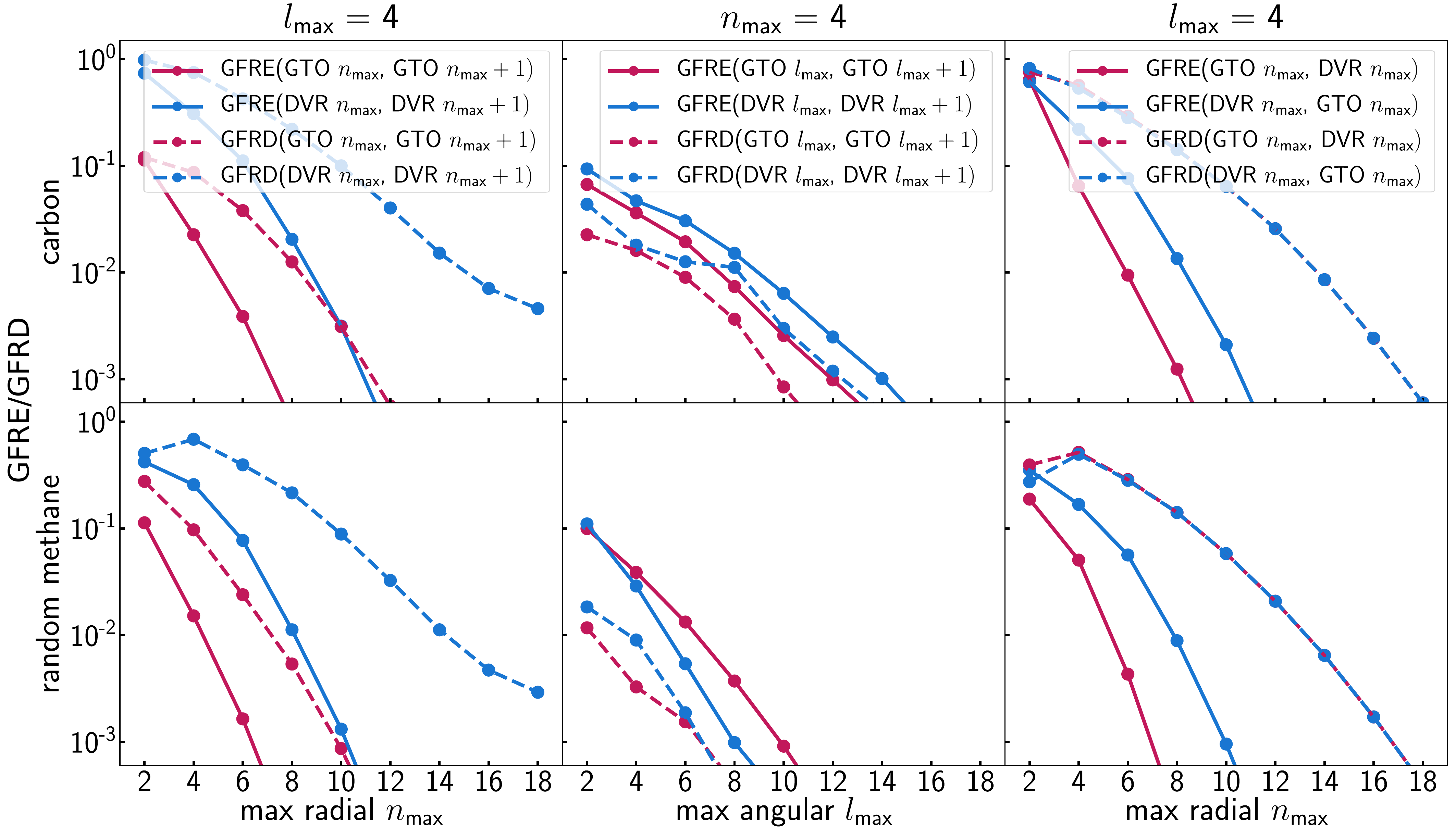}
    \caption{Comparison of the GFRE and GFRD for increasing numbers of radial (left, with fixed $\lmax=4$) and angular (middle, with fixed $\nmax=4$) basis functions.  On the right, an explicit comparison of the two basis sets in terms of $\GFRE(\text{GTO},\text{DVR})$, $\GFRE(\text{DVR},\text{GTO})$ and the corresponding measures of distortion.}
    \label{fig:soap-convergence}
\end{figure*}

\section{Comparing atom-centred representations}

Atom-centred representations that are based on a symmetrized expansion of the atom density constitute one of the most successful and widely adopted classes of features for atomistic machine learning~\cite{will+19jcp,behl-parr07prl,bart+10prl,bart+13prb,thom+15jcp,drau19prb}.
The construction begins by describing a structure $A$ in terms of a sum of localized functions $g(\bx)\equiv\rep<\bx||g>$ (e.g. a Gaussian with variance $\sigmag/2$)  centred on the atom positions $\br_i$
\begin{equation}
\label{eq:density}
\rep<\bx||A; \rho> = \sum_i \rep<\bx - \br_i||g>.
\end{equation}
Symmetrizing over translations and rotations leads to a description of the structure in terms of a sum of \emph{environment} features
\begin{equation}
\rep<\bx_1; \ldots \bx_{\nu}||A; \frho_i^\nu>
\end{equation}
that describe $\nu$-point correlations of the density centred on atom $i$ (effectively corresponding to a $(\nu+1)$-body correlation function in the sense used e.g. in statistical mechanics of liquids).
Different values of $\nu$ correspond to conceptually distinct descriptions of the system -- higher body order terms being more complicated, but potentially more information-rich -- while different discretizations of the abstract vectors on a basis  are a matter of computational convenience and affect the computational cost of different approaches~\cite{zuo+20jpcl}, but their descriptive power should become equivalent in the limit of a complete basis set.
We demonstrate the use of the GFRE, LFRE and GFRD to assess with quantifiable measures the effect of some of the different choices one can make when designing a representation.

\subsection{ SOAP and symmetry functions }
\label{sub:hypers}

We begin by considering two practical realizations of atom-centred symmetrized features of order $\nu=2$: smooth overlap of atomic positions (SOAP) features~\cite{bart+13prb}, as implemented in librascal\cite{LIBRASCAL},  and Behler-Parrinello symmetry functions (BPSF)\cite{behl11jcp} as implemented in the n2p2 package\cite{singraber2019parallel}.
In the SOAP representation the atom-centred density is written as a sum of Gaussians with finite width $\sigmag$, and the density is expanded in a basis that is a product of spherical harmonics $Y^l_m(\bxhat)\equiv\rep<\bxhat||lm>$ and a radial basis $R_n(x)\equiv\rep<x||n>$,
\begin{equation}
\rep<nlm||A; \rho_i> = \sum_j \int \D{\bx}
\rep<n||x> \rep<lm||\bxhat>
\rep<\bx-\br_{ji}||g>, \label{eq:nlm-rho}
\end{equation}
where $\br_{ji}=\br_j-\br_i$.
We consider two different basis sets here, Gaussian-type orbitals (GTO)
\begin{subequations}
\begin{gather}
\rep<r||n; \text{GTO}> = N_n r^{n}\exp(-b_nr^2),\\
\text{with }N_n=\frac2{\sigma_n^{2n+3}\Gamma((n+3)/2)},\\
b_n = 1/(2\sigma_n),\quad \sigma_n = r_c\,\textrm{max}(\sqrt{n},1)/\nmax,
\end{gather}
\end{subequations}
that are orthogonalized with respect to each other,
and a discrete variable representation (DVR) basis
\begin{equation}
\rep<r||n; \text{DVR}>= \sqrt{w_n}\delta(r-r_n)
\end{equation}
where $r_n$ are Gaussian quadrature points and $w_n$ their corresponding weights.
For both bases, the integral~\eqref{eq:nlm-rho} can be evaluated analytically, and the density coefficient computed as a sum over the neighbours of the $i$-th atom. The invariant SOAP features are then computed as a contraction of the density coefficients,
\begin{equation}
\label{eq:SOAP}
\rep<nn'l||\frho_i^2> =
\sum_m  \frac{{(-1)}^m }{\sqrt{2l+1}}  \rep<nlm||\frho_i> \rep<n'l(-m)||\frho_i>
\end{equation}

\begin{figure*}
    \includegraphics[width=0.9\linewidth]{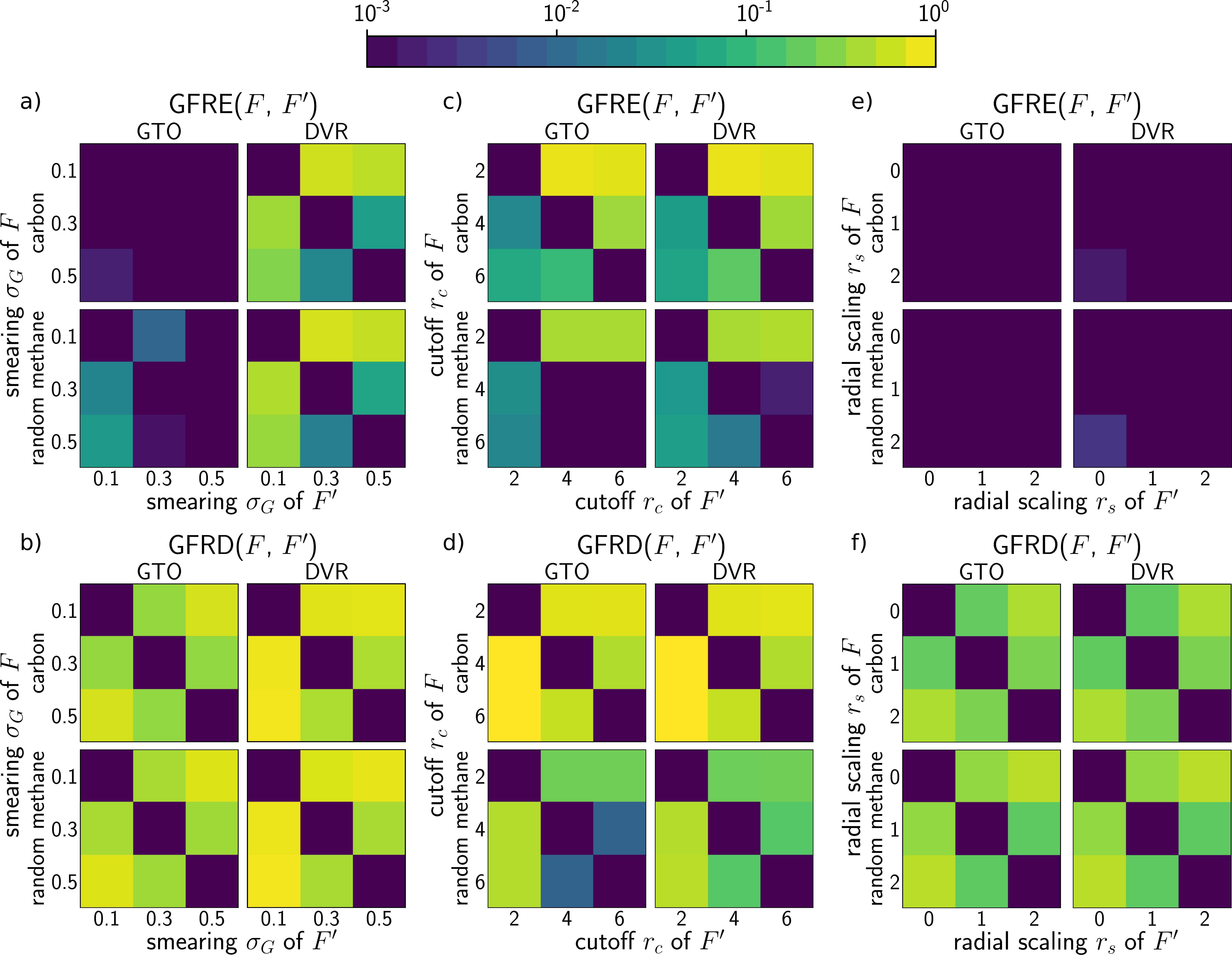}
    \caption{Comparison of the GFRE (top) and GFRD (bottom) a),b) for different smearing  $\sigma_G$ ($r_\text{c}=4\,$\AA{}) c),d) for different cutoff values ($\sigmag=0.5\,$\AA), and e),f) for different radial scaling exponents ($r_\text{c}=4\,$\AA{}, $\sigmag=0.5\,$\AA). For all comparisons $(\nmax,\lmax) = (10,6)$ were used. The feature specified by the row is used to reconstruct the feature specified by the column.}
    \label{fig:soap-sigma-radial-body}
\end{figure*}

Even though they can be seen as a projection on an appropriate basis of the symmetrized atom density that underlies SOAP~\cite{will+19jcp}, Behler-Parrinello symmetry functions (BPSF) are usually computed as a sum over tuples of neighbouring atoms of functions of interatomic angles and distances. Among the many functional forms that have been proposed~\cite{behl11pccp} we consider the two-body functions
\begin{equation}
    G^{(2)}_i = \sum_j e^{-\eta (r_{ji} - R_s)^2} \cdot f_c (r_{ji}), \label{eq:g2}
\end{equation}
and the three-body functions
\begin{multline}
    G^{(3)}_i = 2^{1-\zeta} \sum_j \sum_{k \neq j } (1 + \lambda \cdot \cos \hat{\br}_{ij}\cdot\hat{\br}_{ik})^{\zeta} \cdot \\
    e^{-\eta (r_{ji}^2 + r_{ik}^2 + r_{jk}^2)} \cdot f_c(r_{ji}) f_c(r_{ik}) f_c(r_{jk}), \label{eq:g3}
\end{multline}
where $f_c$ is a cutoff function, and $\eta,\zeta,\lambda,R_s$ are parameters that define the shape of each BPSF. We generate systematically groups of symmetry functions of different size by varying the values of these parameters following the prescriptions discussed in Ref.~\citenum{imba+18jcp}. The list of values for the BPSF parameters we used are supplied in supplementary information.

\paragraph*{GTO and DVR radial basis.} We start by considering the convergence of the SOAP representation with different choices of radial basis. Figure~\ref{fig:soap-convergence} demonstrates the convergence  with the number of radial functions $\nmax$ and angular momentum channels $\lmax$ (in a Cauchy sense, i.e. comparing results for successive increments of these parameters). Overall, the GTO basis converges faster than DVR for most cases, both in terms of GFRE and GFRD.
The slower radial convergence of the GFRD indicates that even as the discretization approaches convergence, the changing position of peaks and nodes of the basis functions gives different emphasis to interatomic correlations over different ranges. This is consistent with the observation that, particularly for small $(\nmax, \lmax)$, regression accuracy depends on the number of basis functions in a way that is not necessarily monotonic.
When considering the convergence of the angular component $\lmax$, GTO and DVR show nearly identical error decay, indicating that the convergence of the radial and angular basis are largely independent of each other.

The faster convergence of the GTO basis suggests that, for a given $\nmax$, a representation expanded on this basis should contain a greater amount of information on the structure. This is reflected in the direct comparison of the two bases, $\GFRE(\text{GTO}\,\nmax,\text{DVR}\,\nmax)<\GFRE(\text{DVR}\,\nmax,\text{GTO}\,\nmax)$ for small $\nmax$.
When both basis set have converged, they become essentially equivalent. Since the two representations are related to each other by a unitary transformation,  $\GFRD(\text{GTO}\,\nmax,\text{DVR}\,\nmax)\rightarrow 0$ as $\nmax\rightarrow\infty$.

\paragraph*{Gaussian smearing.}
The Gaussian smearing used in SOAP features works as a parameter controlling the balance between local resolution and the smoothness of the mapping between Cartesian coordinates and symmetrized density features. A small $\sigmag$ value can identify minute changes more accurately, but a too small value for $\sigmag$ can lead to ill-conditioned regression, as the features associated with different structures show little overlap with each other.
In fact, there is a tight interplay between the density smearing, the choice of the basis set, and the regularization of a regression model.
As seen in Fig.~\ref{fig:soap-sigma-radial-body}(a,b), in the case of the smooth GTO basis set there is relativley little reconstruction error, and in general smaller $\sigmag$ values give a better reconstruction of large-$\sigmag$ features than vice versa. The opposite is true for the $\delta$-like DVR basis: the GFRE for DVR is larger than in the case of GTO, and it is harder to reconstruct large-$\sigmag$ features from their sharp-Gaussian counterparts than vice versa. It should be also added that, without an automatic choice of regularization, results depend greatly on the way the feature mapping is executed. In particular, sharp-to-smooth mapping can lead to major overfitting problems, with $\GFRE$ becoming much larger than one for the test set.
Even in cases where the $\GFRE$ is small, the feature space distortion is large, which highlights the fact that the Gaussian smearing changes significantly the emphasis given to different structural correlations, and can therefore affect the accuracy of regression models.

\paragraph*{Radial cutoff and scaling.}

One of the most important hyperparameters when defining an atom-centred representation is the cutoff distance, which restricts the contributions to the density to the atoms with $r_{ji}<r_{\text{c}}$.
Fig.~\ref{fig:soap-sigma-radial-body}(c,d) shows that the GFRE captures the loss of information associated with an aggressive truncation of the environment, with very similar behavior between GTO and DVR bases.
The figure also reflects specific features of the different data sets: for instance, $\GFRE(r_\text{c}=4\,\text{\AA},r_\text{c}=6\,\text{\AA})$ is close to zero for the random methane data set, because there are no structures where atoms are farther than $4\,$\AA{} from the centre of the environment. $\GFRE>0$ also when mapping long-cutoff features to short-range features, although the reconstruction error is much smaller than in the opposite direction.
This indicates the need for an increase in $\nmax$ to fully describe the structure of an environment when using a large value of $r_\text{c}$, which is consistent with the greater amount of information encoded within a larger environment.
The GFRD plot also underscores the strong impact of the choice of $r_\text{c}$ on the emphasis that is given to different parts of the atom-density correlations.
This effect explains the strong dependency of regression performance on $r_\text{c}$, and the success of multi-scale models that combine features built on different lengthscales~\cite{bart+17sa}.
A similar modulation of the contributions from different radial distances can be achieved by scaling the neighbour contribution to the atom-centred density by a decaying function, e.g. $1/(1+(r_{ji}/r_0)^s)$. This approach has proven to be very effective in fine-tuning the performance of regression models using density-based features~\cite{fabe+17jctc,will+18pccp,paru+18ncomm}.
As shown in Fig.~\ref{fig:soap-sigma-radial-body}(e,f), this is an example of a transformation of the feature space that entails essentially no information loss -- resulting in a very small GFRE between different values of the scaling exponent $s$. However, it does result in substantial $\GFRD$, providing additional evidence of how the emphasis given by a set of features to different inter-atomic correlations can affect regression performance even if it does not remove altogether pieces of structural information.

\begin{figure}
    \includegraphics[width=0.9\linewidth]{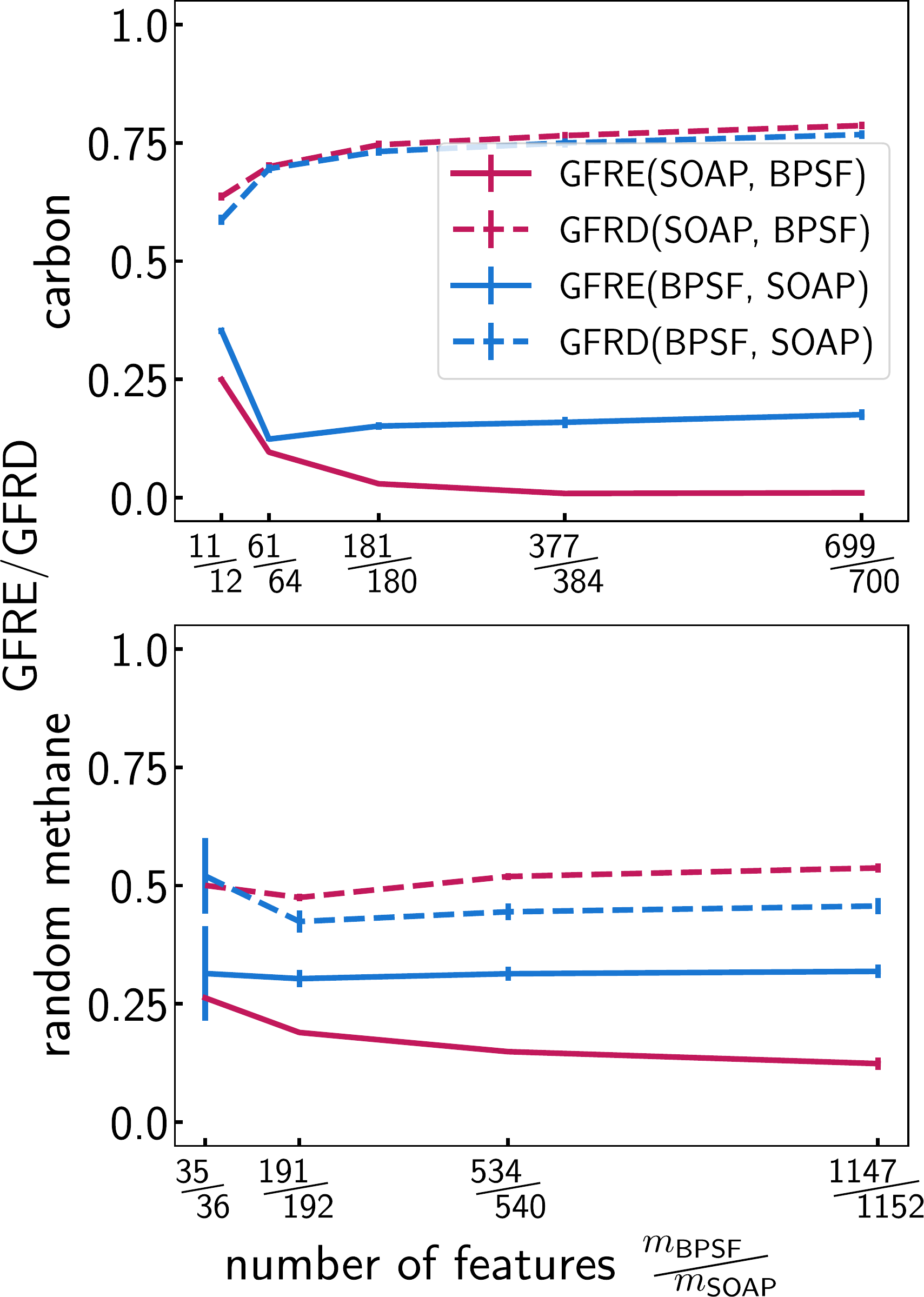}
    \caption{Comparison of the GFRE and the GFRD between SOAP(GTO) and BPSF features with systematically-increasing sizes of the feature vectors. BPSF features are generated by varying over a grid the hyperparameters entering the definitions of $G^{(2)}$ and $G^{(3)}$, following Ref.~\citenum{imba+18jcp}. SOAP expansion truncation parameters $(\nmax,\lmax)$ are adjusted to approximately match the number of BPSF features.}
    \label{fig:soap_behler_parinello}
\end{figure}

\begin{figure}
    \includegraphics[width=0.9\linewidth]{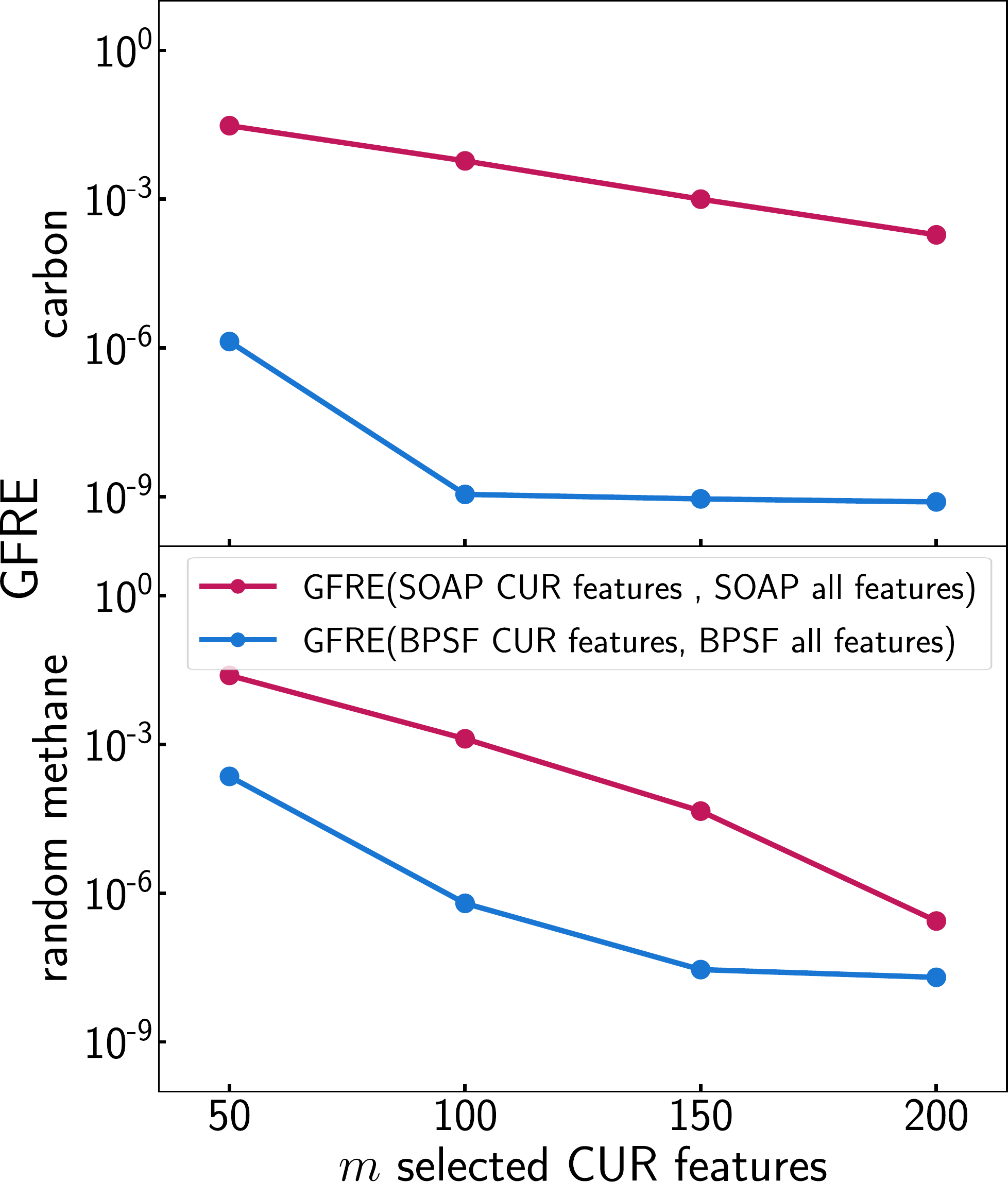}
    \caption{Convergence of a CUR approximation of the full SOAP/BPSF feature vectors (the largest size considered in Fig.~\ref{fig:soap_behler_parinello} )  with number of retained features.}
    \label{fig:soap_behler_parinello-convergence}
\end{figure}

\paragraph*{Behler-Parinello symmetry functions.}

BPSF can be seen as projections of the same, abstract symmetrized density features that underlies the construction of SOAP features.  While the latter representation is usually implemented using an orthogonal set of basis functions, BPSFs are non-orthogonal, and are usually chosen based on a careful analysis of the inter-atomic correlations that are relevant for a given system~\cite{behl11jcp,jose+12jcp,behl15ijqc}, or selected automatically out of a large pool of candidates~\cite{imba+18jcp}.
Fig.~\ref{fig:soap_behler_parinello} shows clearly that an orthogonal basis set provides a more effective strategy to converge a representation than the grid-based enumeration of the non-linear hyperparameters of non-orthogonal basis functions. GFRE(SOAP, BPSF) $<$ GFRE(BPSF, SOAP) for all feature set sizes and both data sets. As usual, we remark that zero reconstruction error does not imply equivalence for regression purpose: the GFRD remains very high even for the largest feature set sizes.

Given that, in real scenarios, one would usually combine systematic enumeration of BPSF features with an automatic selection method\cite{imba+18jcp}, we also use the feature reconstruction framework to investigate the convergence of the automatic screening procedure, i.e. the error in reconstructing the full vector based on the first $m$ features chosen with a CUR decomposition-based procedure\cite{mahoney2009cur, imba+18jcp}.
Figure~\ref{fig:soap_behler_parinello-convergence} shows that a few dozens CUR-selected features allow to almost-perfectly reconstruct the full feature vector. The convergence is particularly fast for BPSF, where $m=50$ leads to a minuscule GFRE, indicating that the non-orthogonal features are highly redundant, and explaining the saturation in model performance that was observed in Ref.~\citenum{imba+18jcp}.

\begin{table*}[bthp]
\revaddfig{
\begin{tabular}{c|c|c|c||c|c}
\hline\hline
    $(n_\text{max},l_\text{max})$ & $m_\text{SOAP}$ & $t_\text{SOAP-GTO}$ / s&  $t_\text{SOAP-DVR}$ / s & $m_\text{BPSF}$ & $t_\text{BPSF}$ / s\\
    \hline
    (2,2) &  12 &  8.08 $\pm$ 0.30 &  7.95 $\pm$ 0.03 &  11 &   3.90 $\pm$ 0.14 \\
    (4,3) &  64 & 10.58 $\pm$ 0.03 &  9.57 $\pm$ 0.02 &  61 &  10.43 $\pm$ 0.22 \\
    (6,4) & 180 & 13.87 $\pm$ 0.03 & 11.87 $\pm$ 0.14 & 181 &  30.20 $\pm$ 0.41 \\
    (8,5) & 384 & 17.58 $\pm$ 0.05 & 14.50 $\pm$ 0.02 & 377 &  66.55 $\pm$ 0.55 \\
   (10,6) & 700 & 23.02 $\pm$ 0.04 & 18.57 $\pm$ 0.03 & 699 & 124.60 $\pm$ 0.78 \\
  \hline\hline
\end{tabular}
}
\caption{\revadd{Timings in seconds for the evaluation of SOAP features using GTO ($t_\text{SOAP-GTO}$) and DVR ($t_\text{SOAP-DVR}$) as radial basis, and of BPSF $(t_\text{BPSF}$), using $r_c=4$\,\AA{}, on the entire carbon dataset. The SOAP discretization parameters are chosen to approximately match the number of features $m_\text{SOAP}$ and $m_\text{BPSF}$.
The measurements have been conducted on a single  Intel(R) Xeon(R) CPU E3-1245 v6 @ 3.70GHz core, using librascal~\cite{LIBRASCAL} for SOAP features and n2p2~\cite{N2P2} for BPSF.}}
\label{table:timings}
\end{table*}

\revadd{\paragraph*{Computational cost}
In this work we focus on the comparison of  different kinds of features in terms of their information content, without commenting on the computational overhead associated with their evaluation, or the application of non-linear transformations.
Computational cost depends on implementation choices, and can be optimized for usage patterns that differ from those that we apply here.
However, the effort associated with the evaluation of a model plays an important role in determining its ultimate usability. To provide some context for our experiments, we report in Table~\ref{table:timings} the timings for the evaluation of SOAP and BPSF features with the same parameters used in this Section.
These timings show clearly that the computational savings afforded by the use of a DVR basis are not sufficient to offset the reduced information content with respect to the GTO basis.
When comparing BPSF and SOAP, the clearest difference is that the cost of evaluating the former scales linearly with the number of features, while the cost of evaluating SOAP features is sublinear since it is dominated by the calculation of the density expansion coefficients $\bra{nlm}\ket{\rho_i}$ and is therefore sublinear with respect to the invariants.
This difference in scaling is due to the different mechanism for evaluating 3-body terms, that scales quadratically with the number of neighbours for BPSF, and only linearly for SOAP, underscoring how a careful selection of the most relevant features is very important in the context of a BPSF framework.
}

\begin{figure}
    \includegraphics[width=1.0\linewidth]{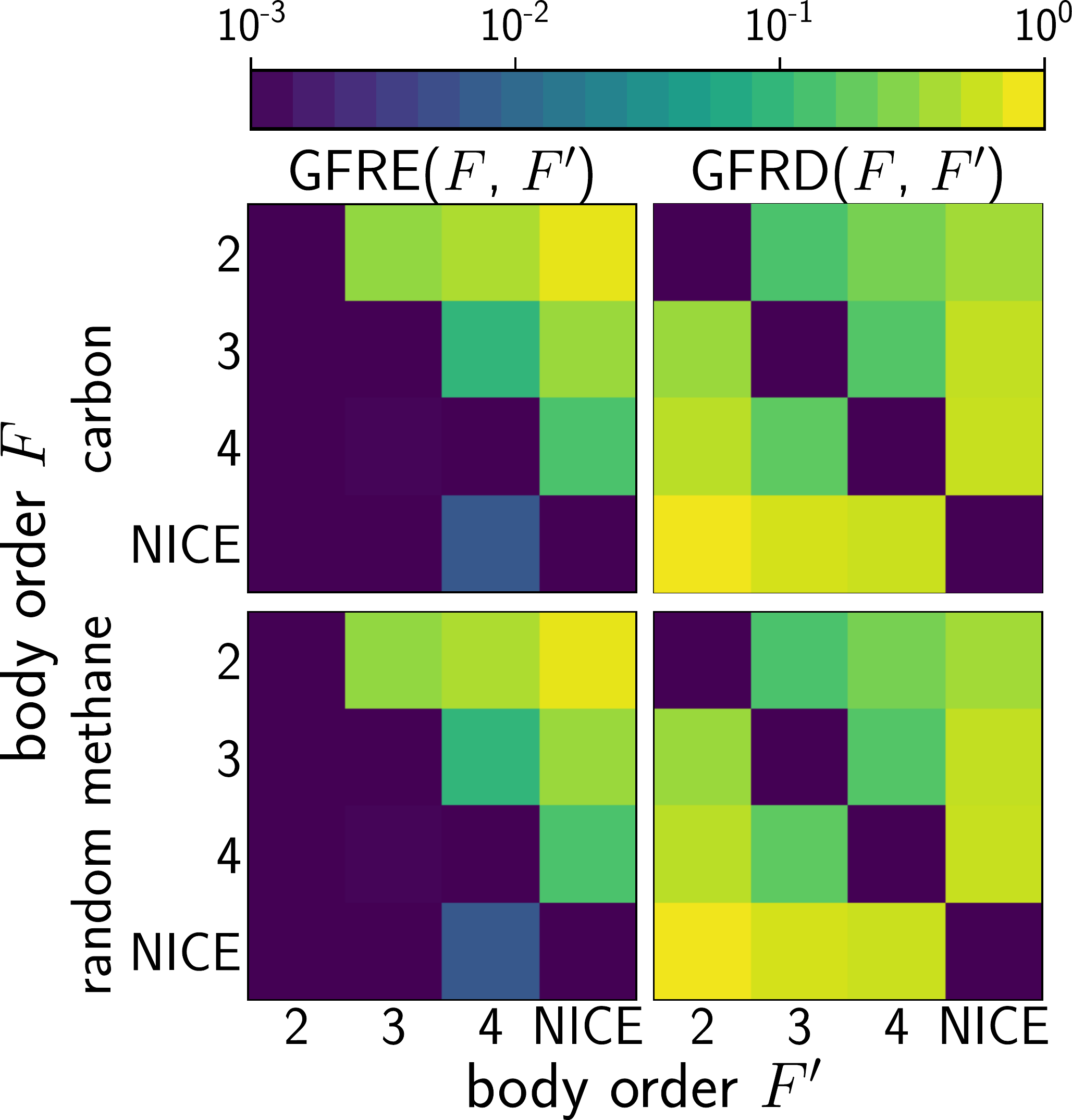}
    \caption{GFRE and GFRD body order comparison using GTO as radial basis function, $r_c= 4\,$\AA{}, $\sigma_G=0.5\,$\AA{} and $(n_\text{max},l_\text{max})=(6,4)$. NICE features were computed keeping the top 400 equivariant components at each level of the body-order iteration, and keeping invariant components up to $\nu=4$.  }
    \label{fig:soap-bodyorder}
\end{figure}

\subsection{ Body order feature truncation. }

The examples in Section~\ref{sub:hypers} demonstrate the impact of implementation details and hyperparameters choices on the information content of features that are all equivalent to a three-body correlation of the atom density. A more substantial issue is connected to the use of representations based on different $\nu$-body correlations of a decorated atom density, which is equivalent to the pair correlation function (2-body, $\nu=1$), to the SOAP power spectrum (3-body, $\nu=2$) or to the bispectrum (4-body, $\nu=3$).
Different orders incorporate conceptually distinct kinds of information: when used in linear regression, different density correlation orders correspond to a body-order expansion of the target property~\cite{shap16mms,glie+18prb,will+19jcp,drau19prb,jinn+20jcp}, and the link between the convergence of the body-order expansion and the injectivity of the structure-feature map is an open problem, with known counter-examples showing that low values of $\nu$ are insufficient to achieve a complete representation of an atomic environment~\cite{pozd+20prl}.

\begin{figure}
    \includegraphics[width=0.85\linewidth]{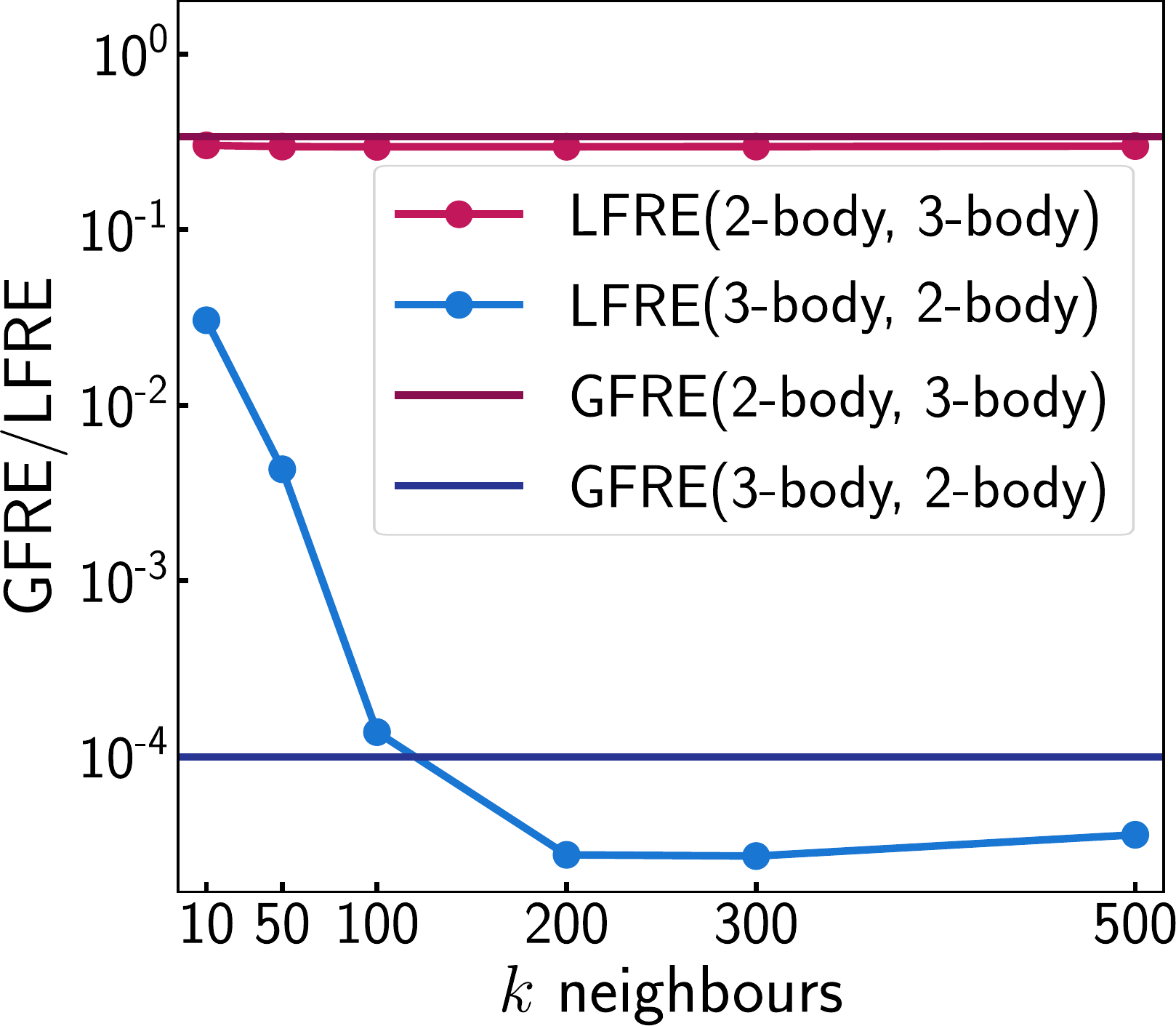}
    \caption{Convergence of the LFRE between 2 and 3-body density correlation features (using GTOs as radial basis, $r_c= 4\,$\AA{}, $\sigma_G=0.5\,$\AA{} and $(n_\text{max},l_\text{max})=(6,4)$) with increasing number of neighbours.}
    \label{fig:soap-lfre-convergence}
\end{figure}

\begin{figure}
    \includegraphics[scale=0.42]{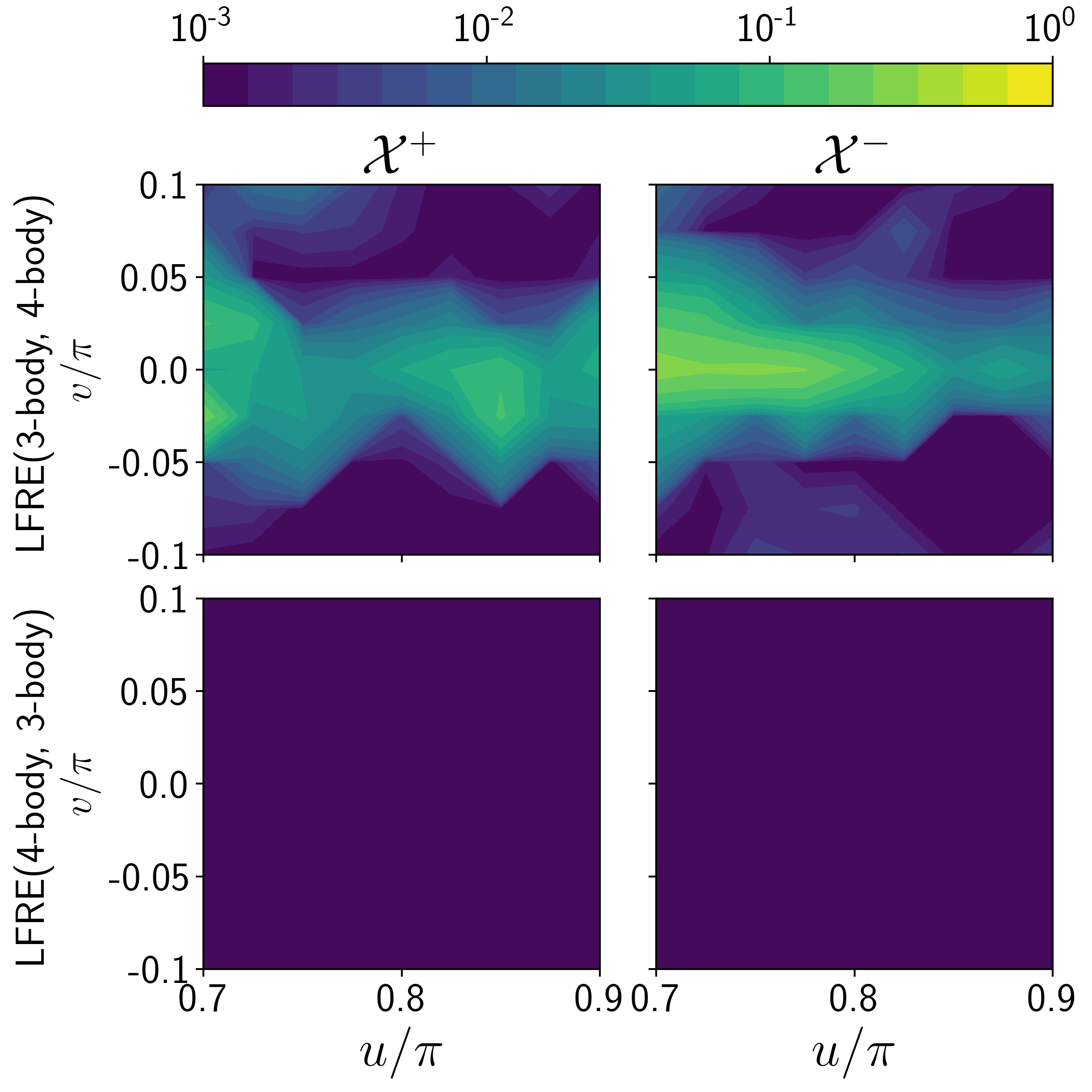}
\caption{Pointwise LFRE for the structures from the degenerate methane dataset as a function of the structural coordinates $(u,v)$ for $(\nmax,\lmax) = (6,4)$ and $k=15$ neighbours.}
    \label{fig:soap_degenerated_manifold_lfre}
\end{figure}

Fig.~\ref{fig:soap-bodyorder} shows that high-order features cannot be recovered as linear functions of lower-order features, while an approximate (if not complete) reconstruction of lower-$\nu$ components based on high-$\nu$ components is possible. Reconstructing features of different order entails a large amount of distortion, with the GFRD approaching one in most cases.
We also include in the comparison features obtained with the recently-developed $N$-body iterative contraction of equivariants (NICE) framework, that identifies the most important features for each $\nu$ value, and uses them to compute  $(\nu+1)$-order features~\cite{niga+20jcp}. Keeping 400 features for each body order is sufficient to achieve perfect reconstruction of 2 and 3-body features, but not for the 4-body (bispectrum) term, which cannot be reconstructed fully with 400 NICE features. Considering however that $\GFRE(\text{NICE},\nu=3)\ll \GFRE(\nu=3, \text{NICE})$, one can infer that the of information loss associated with truncating the body order expansion is more severe than when restricting the number of 4-body features.

The comparison of features of different order can also be used to elucidate the role of the (non-)linearity of the mapping between feature spaces. Figure~\ref{fig:soap-lfre-convergence} compares global and local feature reconstruction errors between 2 and 3-body density correlation features, for the random $\ce{CH4}$ data set. In the case of the low-to-high body order reconstruction, the LFRE is only marginally lower than its global counterpart, indicating that the large $\GFRE(\nu=1,\nu=2)$ is a consequence of lower information content and not only of the linear nature of the map. The reverse case is also revealing: for small $k$-neighbourhood sizes, $\LFRE(\nu=2,\nu=1)>\GFRE(\nu=2,\nu=1)$, because the small number of neighbours included in the model reduce the accuracy of the feature reconstruction map. When the number of neighbours approaches the intrinsic dimensionality of the $\nu=2$ features, instead, $\LFRE<\GFRE$ -- because the reconstruction is based on a locally-linear map that can approximate a non-linear relationship between features. As $k$ approaches the full train set size, the LFRE approaches the GFRE, as the locality of the mapping is lost.

The LFRE also makes it possible to identify regions of phase space for which the construction of a mapping between feature spaces is difficult or impossible.
Consider the case of the degenerate manifold discussed in Ref.~\citenum{pozd+20prl}. The dataset includes two sets of \ce{CH4} environments, and those parameterised by $v=0$ cannot be distinguished from each other using 3-body ($\nu=2$) features.
Fig.~\ref{fig:soap_degenerated_manifold_lfre} shows the LFRE for each point along the two manifolds. When trying to reconstruct 3-body features using as inputs 4-body features (that take different values for the two manifolds) the LFRE is essentially zero. When using the 3-body features as inputs, instead, one observes a very large error for points along the degenerate line, while points that are farther along the manifold can be reconstructed well. This example demonstrates the use of the LFRE to identify regions of feature space for which a simple, low-body-order representation is insufficient to fully characterize the structure of an environment, and can be used as a more stringent, explicit test of the presence of degeneracies than the comparison of pointwise distances discussed in Ref.~\citenum{pozd+20prl}.

\subsection{Kernel-induced feature spaces}

\begin{figure}
    \includegraphics[width=0.9\linewidth]{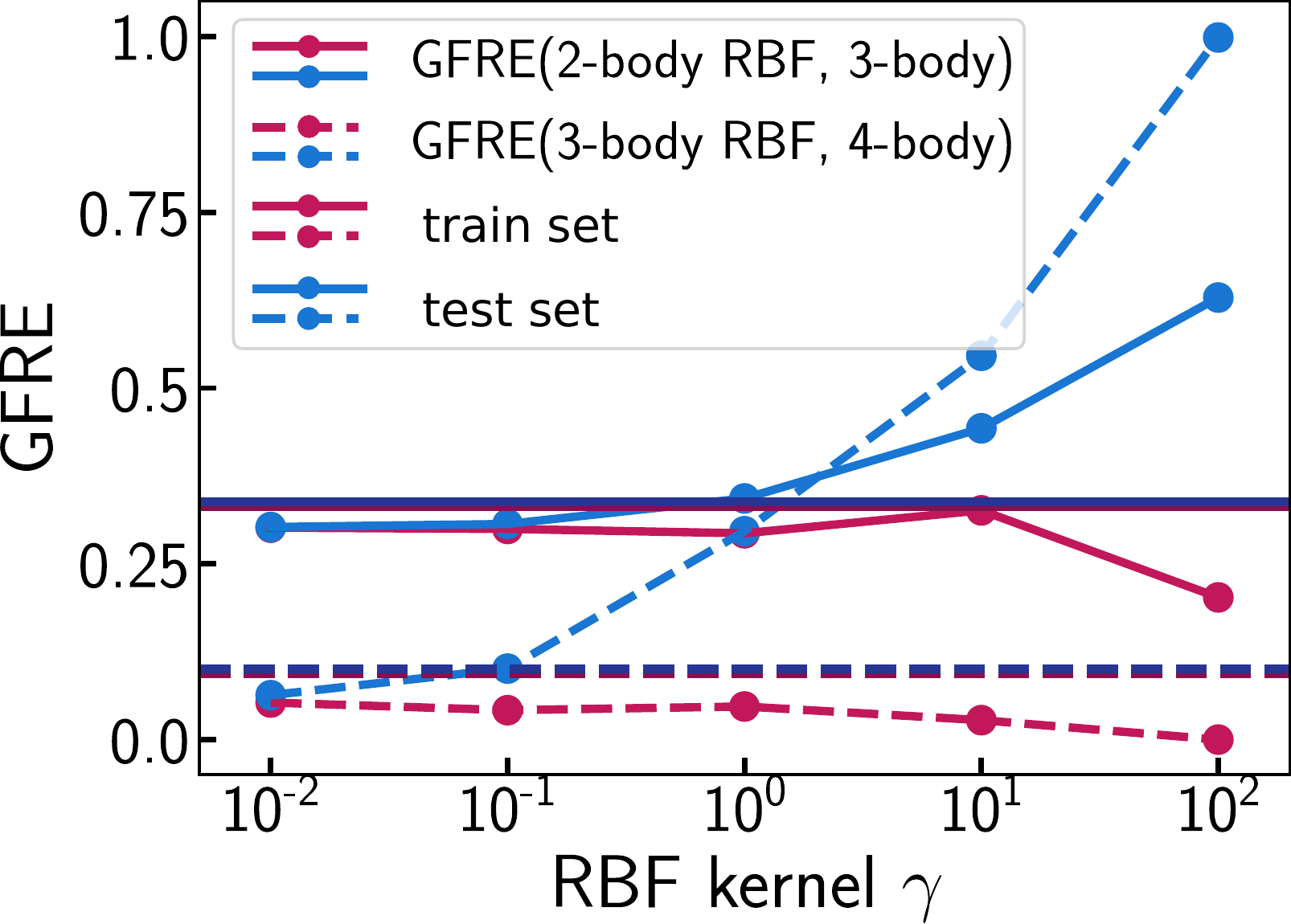}
    \caption{GFRE on the random methane dataset curves as a function of the $\gamma$ hyperparameter of $k^{\text{RBF}}_E$.
    Values for train and test sets are plotted separately. The horizontal lines correspond to the GFRE of the linear features.}
    \label{fig:rbf-bo-promote}
\end{figure}

With the exception of the trivial, scalar-product form, a kernel introduces a non-linear transformation of the feature space, potentially allowing to obtain more accurate regression models.
A crucial aspect of kernel methods is the fact that this non-linear transformation gives rise to a linear feature space that is defined by the combination of the kernel and the training samples -- or the active samples in the case of sparse kernel methods.
We can then use our feature-space reconstruction framework to compare quantitatively the linear feature space with the kernel-induced features. We do so using a radial basis function kernel, varying the $\gamma$ parameter. In the $\gamma\rightarrow 0$ limit the RBF kernel becomes roughly linear, and the non-linearity increases with growing $\gamma$. The use of standardized input features means that $\gamma$ is effectively unitless, and also standardize the kernel-induced features.
To reduce noise, given that the kernel matrices are often very ill-conditioned, we only retain the RKHS features that are associated with the largest eigenvalues, preserving those that together contribute to approximately 99\%{} of the variance.
1
Figure~\ref{fig:rbf-bo-promote} plots the GFRE and GFRD for the mapping of linear and RBF features computed for 2 and 3-body density correlations. The non-linear nature of the transformation is apparent in the increase in the GFRE(linear,RBF) for larger values of $\gamma$, for both $\nu=1$ and $\nu=2$. The transformation is not entirely lossless, as evidenced by the fact that the reverse GFRE is also non-zero. The GFRE(RBF,linear) becomes particularly large for very large values of the $\gamma$ parameter. This can be understood from the fact that the decay of the kernel becomes very sharp, and it only provides information about the nearest neighbours of each point -- effectively leading to an ill-conditoned regression problem as we show in more detail below.

\begin{figure}
    \includegraphics[width=0.9\linewidth]{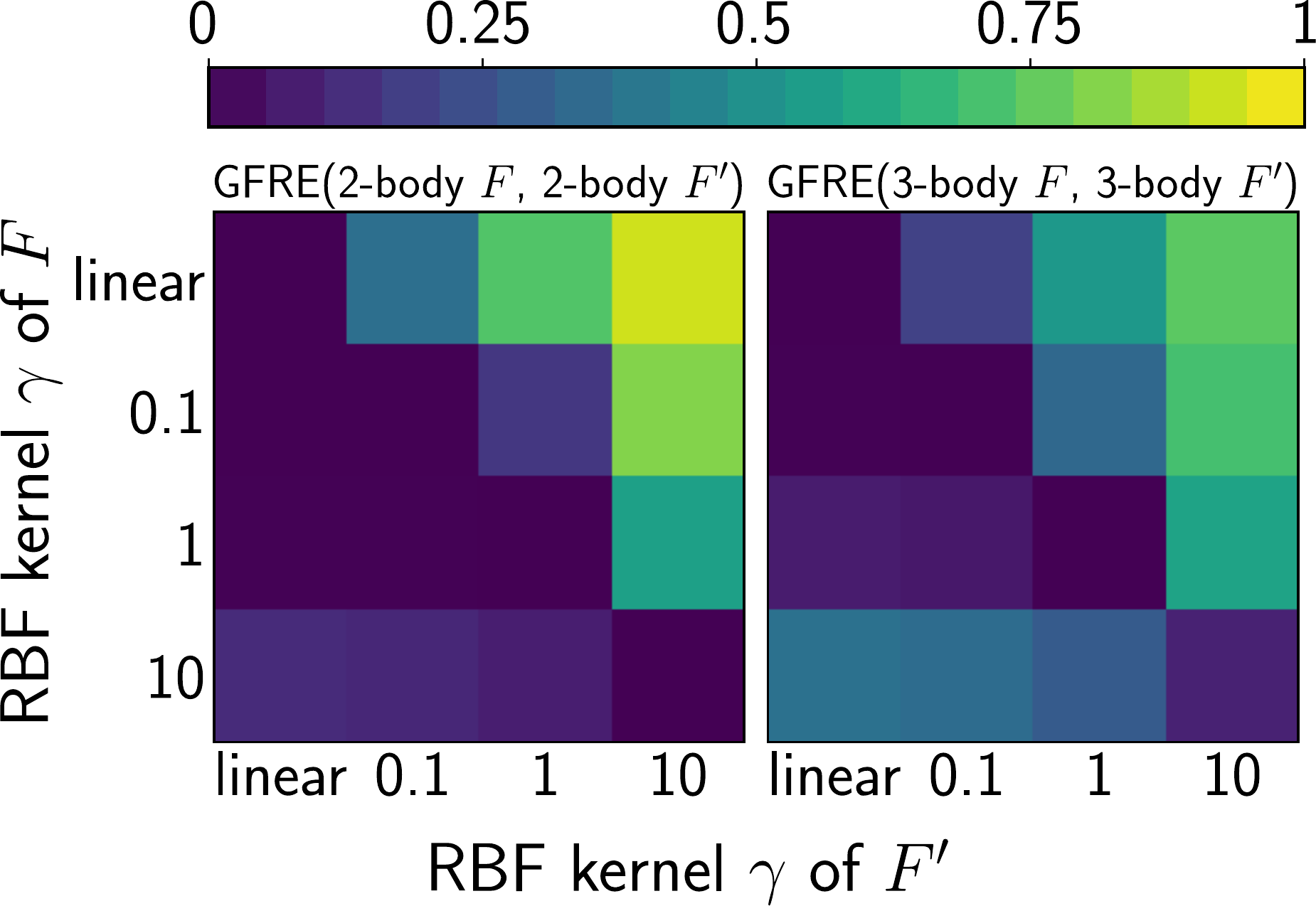}
    \caption{GFRE on the random methane dataset for interconverting the linear 2-body (left) and 3-body (right) feature spaces with those induced by a RBF kernel with different inverse  kernel width $\gamma$.}
    \label{fig:rbf-density}
\end{figure}

\begin{figure*}
\includegraphics[width=0.9\linewidth]{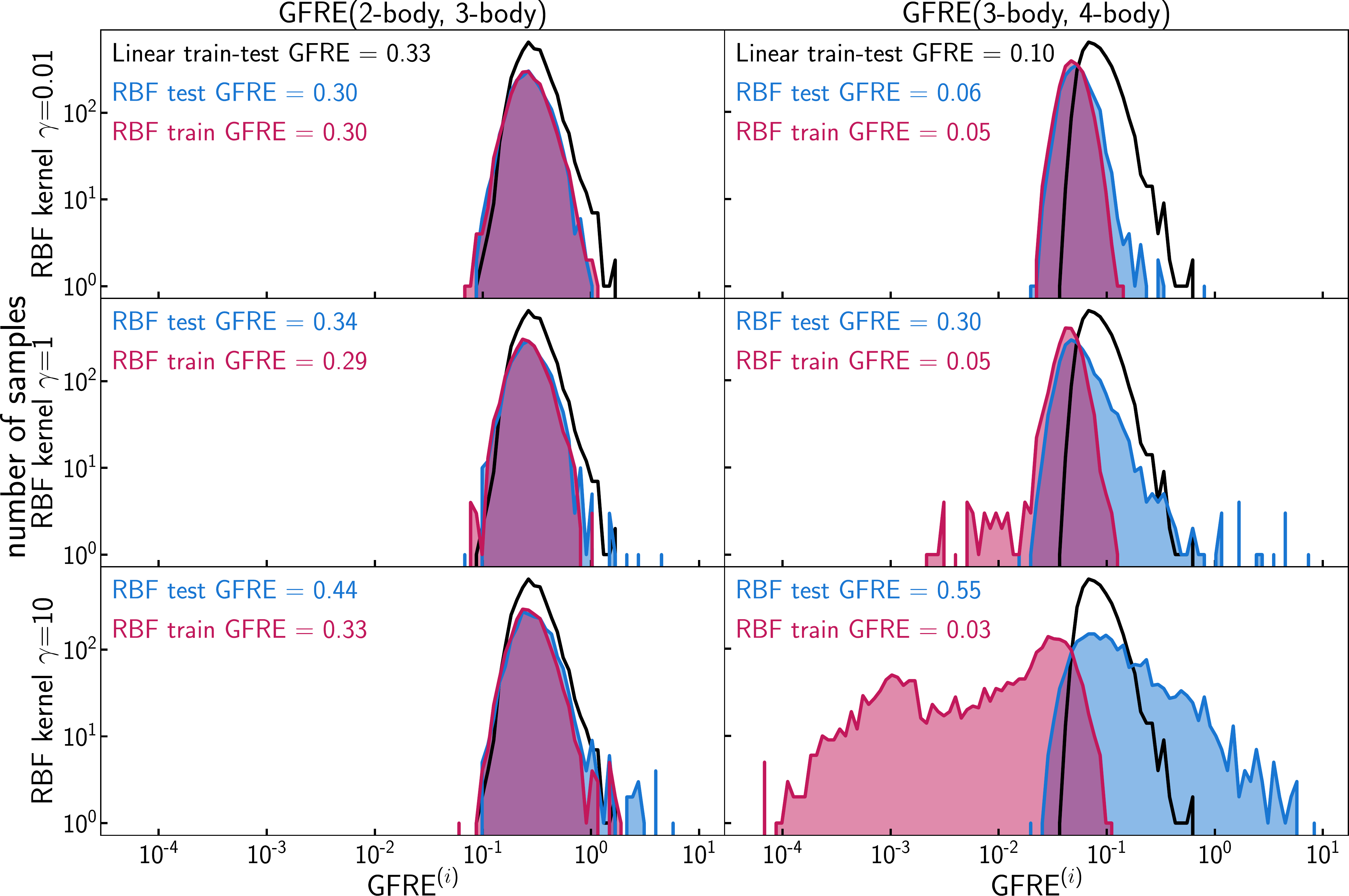}
\caption{Histograms of the pointwise reconstruction error for $2\rightarrow 3$ (left) and $3\rightarrow 4$ (right) body order features, using a RBF kernel with different values of $\gamma$ (top to bottom, $\gamma=$ 0.01, 1.0, 10) to reconstruct the higher body order features. Red curves refer to the train set points, blue curves to the test set, and the black line correspond to the linear train-test set GFRE, that serves as a reference.}
    \label{fig:rbf-body-hists}
\end{figure*}

Having assessed the impact of non-linear kernel features on a single body order representation, we can then investigate whether a non-linear transformation helps inferring high-body order correlations from low-body-order features.
This is relevant because the use of non-linear kernels has been proposed~\cite{glie+18prb} (and used in practice for a long time~\cite{bart+10prl,bart+13prb}) as a strategy to describe many-body effects on atomistic properties.
We compute the GFRE for promoting $\nu=1$ (2-body) to $\nu=2$ (3-body) and $\nu=2$ to $\nu=3$ features for different values of the RBF kernel $\gamma$.
In Figure~\ref{fig:rbf-bo-promote} we show these curves for both the usual GFRE definition (that involves a separate test set) and for a prediction carried out on the train set.
These results show that while a non-linear kernel does allow a low-body-order model to discern higher body-order features, it does so in a poorly transferable way: high-$\gamma$ models show much reduced GFRE for train-set predictions, but lead to a degradation in the feature reconstruction for the test set.
Only low-$\gamma$ models show a small improvement in the test-set GFRE compared to an entirely linear mapping. In this regime, the RBF kernel is dominated by the low-exponent components of the Gaussian expansion, vindicating the choice of low-order polynomial kernels, that are used in most of the published SOAP-based potentials.
A better understanding of the effect of a non-linear feature space transformation can be obtained by analyzing the distribution of reconstruction errors for individual samples
\begin{equation}
\GFRE^{(i)}(\CF,\CF') = \norm{\bx'_i - \bx_i\bP{\CF}{\CF'}}.
\label{eq:pointwise_GFRE}
\end{equation}
The histograms for this ``pointwise GFRE'' (Fig.~\ref{fig:rbf-body-hists}) show that increasing the non-linearity of the kernel does indeed allow to reconstruct more accurately a fraction of both the test and the train set. When extrapolating the mapping to points that have not been seen before, however, there is an increasingly large fraction of outliers for which the reconstruction is catastrophically poor.

The pointwise errors are also revealing of the different nature of the $\nu=1\rightarrow \nu=2$ and $\nu=2\rightarrow \nu=3$ cases.
In the former case, the clear lack of information in the 2-body descriptor makes it impossible, even for a highly non-linear kernel, to obtain an accurate reconstruction of higher body-order features.
In the latter case, instead, the train set reconstruction become nearly perfect with large $\gamma$ -- indicating that despite the existence of degenerate manifolds of configurations~\cite{pozd+20prl} it is possible to reconstruct $4$-body features using only $3$-body inputs, for  structures that are not exactly on the degenerate manifold.
However, the increasingly large tail of very high test-set GFRE samples suggests that this mapping is not smooth, and rather unstable. When building a regression model for a property that depends strongly on 4-body terms, this instability may translate into poor extrapolative power for a non-linear model based on 3-body features.

\begin{figure*}
    \includegraphics[width=0.77\linewidth]{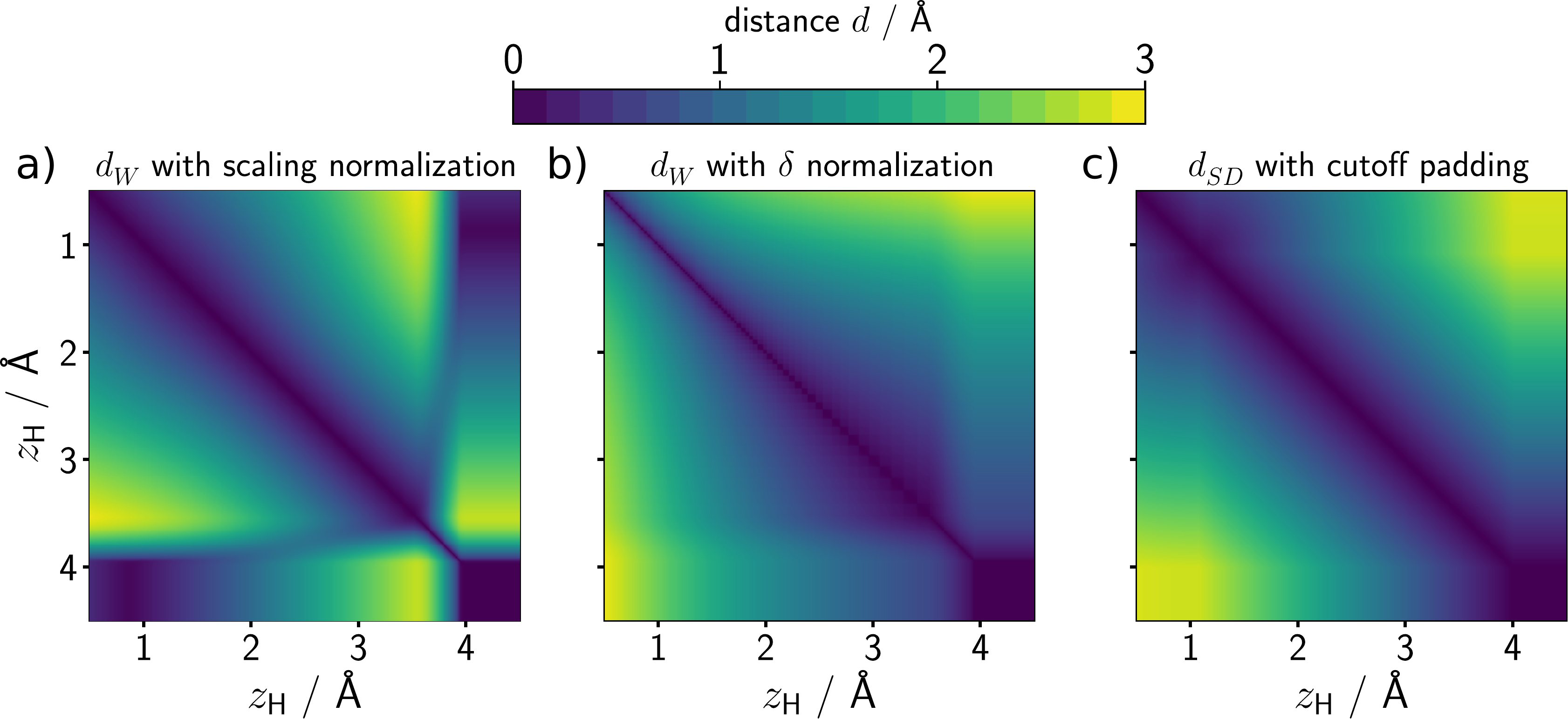}
    \caption{Distance between two \emph{displaced methane} configurations with different values of $\zh$, computed using a Wasserstein distance using (a) scaling normalization; (b) cutoff $\delta$ normalization; (c) Euclidean distance between sorted interatomic distance vectors.}
    \label{fig:wasserstein-distances}
\end{figure*}

\begin{figure}
\includegraphics[width=0.98\linewidth]{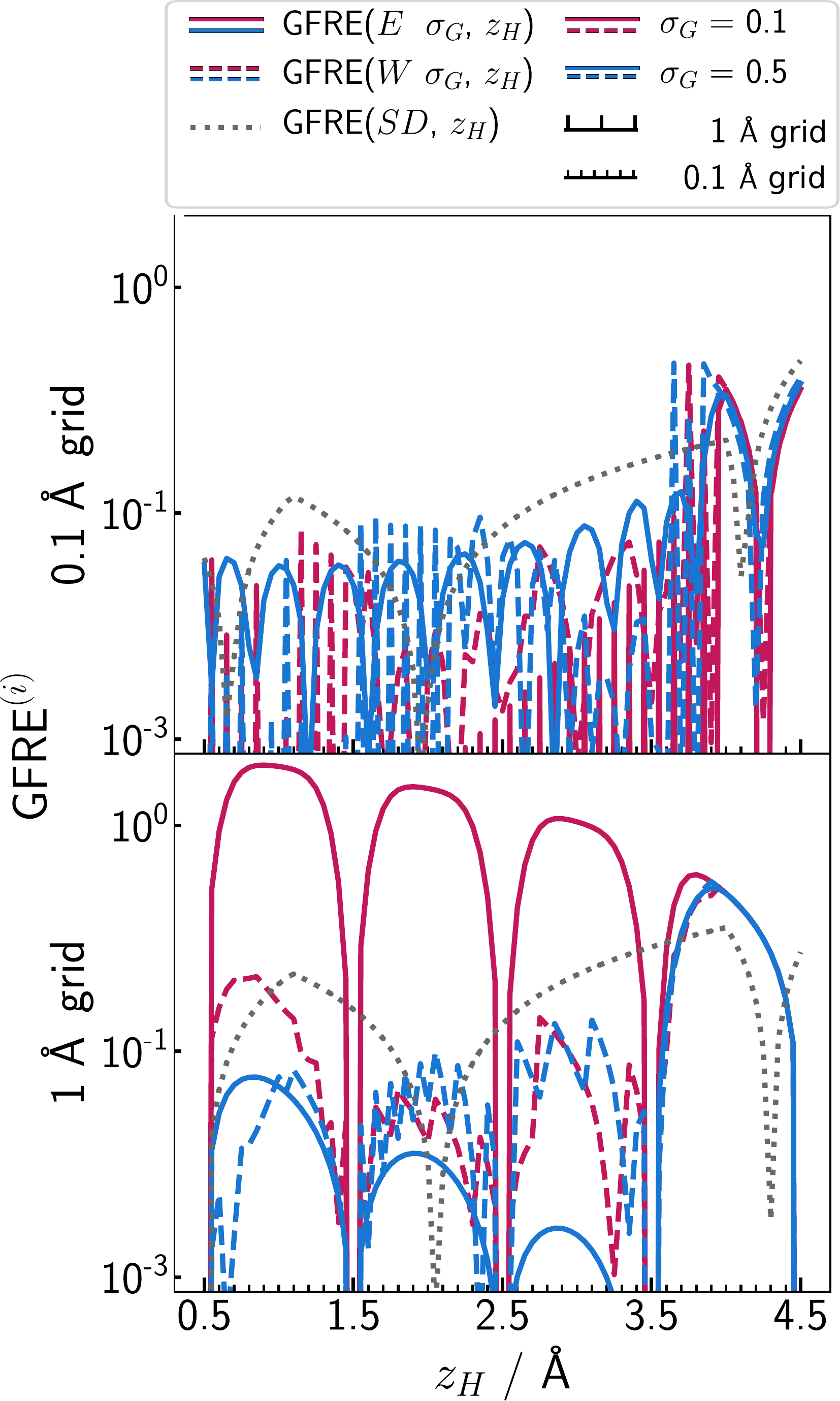}
\caption{
Errors when reproducing the atomic displacement $\zh$ for a fine (top) and coarse (bottom) grid of training points, and different Gaussian $\sigmag$ and metrics. A constant regularization that discards singular values smaller than \revadd{$10^{-3}$} has been applied to all pointwise GFRE calculations.}
\label{fig:wasserstein-grid}
\end{figure}

\subsection{Wasserstein metric}
As an example of the transformation induced by a non-Euclidean metric we consider the effect of using a Wasserstein distance to compare $\nu=1$ density correlation features.
The Wasserstein distance (also known as the Earth Mover Distance, EMD) is defined as the minimum  ``work'' that is needed to transform one probability distribution into another -- with the work defined as the amount of probability density multiplied by the extent of the displacement~\cite{vall74siam,cohe-guib97report,cutu07proc}.
The EMD has been used to define a ``regularized entropy match'' kernel to combine local features into a comparison between structures~\cite{de+16pccp}, to obtain permutation-invariant kernels based on Coulomb matrices\cite{cayl+20mlst}, and  has been shown to be equivalent to the Euclidean distance between vectors of sorted distances~\cite{will+19jcp}.
Here we use the Wasserstein distance to compare two-body ($\nu=1$) features, that can be expressed on a real-space basis and take the form of one-dimensional probability distributions.

The formal definition of the Wasserstein distance of order 2 between two probability distributions $p(r)$ and $p'(r)$ defined on a domain $M$ reads
\begin{equation}
W(p,p')^2 = \hspace{-1em}\inf_{\gamma\in\Gamma (p,p')} \int_{M\times M}\hspace{-0.5em}d(r,r')^2\,\mathrm{d}\gamma(r,r'),
\end{equation}
where $\Gamma(p,p')$ is the set of all joint distributions with marginals $p$ and $p'$.
For 1-dimensional distributions, $W(p,p')$ can be expressed as the 2-norm of the difference between the associated inverse cumulative distribution function (ICDF) $P^{-1}$ of two environments,
$W(p,p')^2=\int_0^1  \left|P^{-1}(s) -{P'}^{-1}(s)\right|^2\D{s}$, with $P(r)=\int_0^r p(r)\D{r}$

In order to express the symmetrized 2-body correlation function as a probability density, we first write it on a real-space basis $\rep<r|$, and evaluate it on 200 Gaussian quadrature points, that we also use to evaluate the CDF and its inverse.
We then proceed to normalize it, so that it can be interpreted as a probability density.
We estimate the integral of the distribution (that effectively counts the number of atoms within the cutoff distance)
\begin{equation}
Z_i = \int_0^{r_c}\rep< r||A;\frho_i^1>\,\mathrm{d}r,
\end{equation}
and the maximum value of the integral over the entire dataset $Z_\CD$.
A simple scaling of the correlation function
\begin{equation}
p_i^{\text{s}}(r) = \frac1{Z_i} \rep< r||A;\frho_i^1>\label{eq:wass-pis}
\end{equation}
distorts the comparison between environments with different numbers of atoms.
To see how, we use the \emph{displaced methane} dataset, in which three atoms in a \ce{CH4} molecule are held fixed in the ideal tetrahedral geometry, at a distance of 1\AA{} from the carbon centre. The fourth atom, aligned along the $z$ axis, is displaced along it, so that each configuration is parameterised by a single coordinate $\zh$.
Figure~\ref{fig:wasserstein-distances}(a) shows the distance computed between pairs of configurations with different $\zh$,  demonstrating the problem with the renormalized probability~\eqref{eq:wass-pis}: $p^{\text{s}}$ loses information on the total number of atoms within the cutoff, and so once the tagged atom moves beyond $r_\text{c}$ the remaining \ce{CH3} environment becomes indistinguishable from an ideal \ce{CH4} geometry.

One can obtain a more physical behavior when atoms enter and leave the cutoff by introducing a $\delta$-like ``sink'' at the cutoff distance, defining
\begin{equation}
p^{\delta}_i(r) = \frac{1}{Z_\CD}\left[\rep< r||\mathcal{A};\frho_i^1> + (Z_\CD-Z_i)\delta(r-r_c)\right].
\end{equation}
Fig.~\ref{fig:wasserstein-distances}b shows that with this choice the Wasserstein metric between $p_i^\delta(r)$ reflects the distance between the moving atoms. With this normalization, in fact, the Wasserstein metric corresponds to a smooth version of the Euclidean metric computed between vectors of sorted interatomic distances~\cite{will+19jcp}, shown in Fig.~\ref{fig:wasserstein-distances}c.
The distortions that can be seen in the comparison between Fig.~\ref{fig:wasserstein-distances}b,c are a consequence of the Gaussian smearing, the smooth cutoff function, and the $\SOthree$ integration that modulates the contribution to $\rep< r||\mathcal{A};\frho_i^1>$ coming from atoms at different distances.

Having defined a meaningful normalization and a probabilistic interpretation of the radial density correlation features, we can investigate how the feature space induced by a Wasserstein metric relates to that induced by an Euclidean distance.
Figure~\ref{fig:wasserstein-grid} shows the error in the reconstruction of $\zh$ for the \emph{displaced methane} dataset when restricting the training set to 0.05\AA{} and 1.0\AA{} spaced grids.
Using a Euclidean distance with a sharp $\sigmag$ leads to a highly non-linear mapping between the displacement coordinate and feature space, and a linear model cannot interpolate accurately between the points of a sparse grid.
A Wasserstein metric, on the other hand, measures the minimal distortion needed to transform one structure into another, and so provides a much more natural interpolation along $\zh$, which is robust even with a sharp density and large spacing between training samples.
It is worth stressing that the sorted distance metric -- which effectively corresponds to the $\delta$ density limit of the Wasserstein metric -- performs rather poorly, and cannot even reproduce the training points.
This is because the mapping between feature space and $\zh$ is not exactly linear, changing slope when $\zh$ crosses 1\AA{} (because the sorting of the vector changes)  and 4\AA{} (because one atom exits the cutoff). The sorted-distances feature space does not have sufficient flexibility to regress this piecewise linear map, as opposed to its smooth Wasserstein counterpart.

\begin{figure}
\includegraphics[width=1\linewidth]{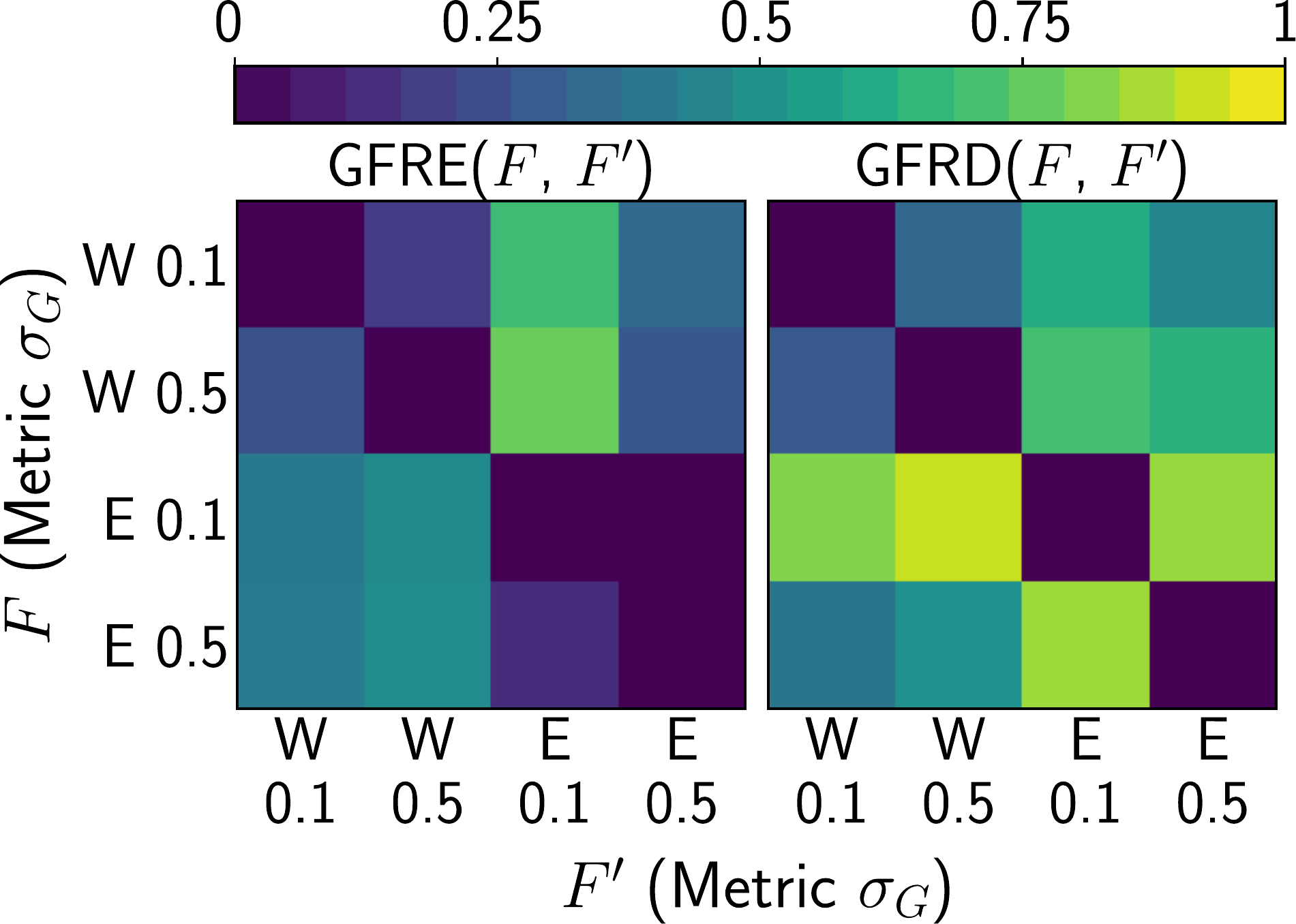}
\caption{Comparison of GFRE and GFRD for the \emph{carbon} dataset, using sharp ($\sigmag=0.1$\AA{}) and smooth ($\sigmag=0.5$\AA{}) radial SOAP features, as well as Euclidean (E) and Wasserstein (W) metrics.}
\label{fig:wasserstein-carbon}
\end{figure}

Having rationalized the behavior of the Wasserstein metric for a toy model, we can test how it compares to the conventional Euclidean metric on a more realistic data set. We consider in particular the AIRSS \emph{carbon} data set, and compare different levels of density smearing as well as Euclidean and Wasserstein metrics.
Figure~\ref{fig:wasserstein-carbon} paints a rather nuanced picture of the relationship between the linear and the Wasserstein-induced feature spaces.
The GFRE is non-zero in both directions, meaning that (in a linear sense) Wassertein and Euclidean features provide complementary types of information.
Smearing of the density has a small effect on the Wasserstein metric, so that both $\GFRE(W(\sigmag=0.1\text{\AA}),W(\sigmag=0.5\text{\AA}))$ and $\GFRD(W(\sigmag=0.1\text{\AA}),W(\sigmag=0.5\text{\AA}))$ are small, whereas for Euclidean features -- as observed in Section~\ref{sub:hypers} -- changing $\sigmag$ induces small information loss, but a large distortion of feature space.
Overall, there is no sign of the pathological behavior seen in Fig.~\ref{fig:wasserstein-grid}, which is an indication that (at least for 2-body features) the \emph{carbon} dataset is sufficiently dense, and that the better interpolative behavior of the EMD does not lead to a more informative feature space.

\section{Conclusion}

Applications of machine learning to atomistic modelling suggest that the featurization that is chosen to represent a molecule or material can be equally or more important than the choice of regression scheme~\cite{fabe+17jctc}.
This has led to the proliferation of approaches to build descriptors, that often differ from each other only in implementation details.
The framework we introduce in this work enables a comparison of alternative choices of representations that does not depend on the target property, and makes it possible to determine objectively which of two features contains more information -- based on a feature-space reconstruction error --  and how much distortion is present in the way they describe the information that is common between the pair -- based on a measure of feature-space distortion.
Even though the framework is linear in nature, it can be generalized to account for non-linear relationships between feature spaces, either by using kernel-induced features, or by decomposing the feature comparison problem into a collection of local mappings.

Using this framework we demonstrate that the choice of basis set can affect substantially the convergence of SOAP features, and that for instance  Gaussian type orbitals are more stable in the limit of small density smearing than the discrete variable representation basis.
In practice the convergence of the representation with the number of basis functions should be considered together with the computational cost of the basis. The analytical expression for GTOs involve special functions that are usually harder to evaluate than those that appear for a DVR basis. This overhead, however, is not sufficient to compensate for the reduced information content, and can be avoided altogether by using a spline approximation for the special functions. In general, computational cost depends substantially on the details of the implementation, and should be assessed in an end-to-end manner as a function of the specific use case.
We also show quantitatively that a systematic orthogonal basis is much more effective in describing the atom density than the heuristic symmetry functions of the Behler-Parrinello kind -- notwithstanding the considerable success that the latter approach has had in the construction of neural-network-based interatomic potentials~\cite{behl16jcp}.

A more systematic difference between atomistic machine-learning frameworks arises from the choice of the order of inter-atomic correlations that underlies the representation.
We show that atom density correlation features of high body order make it possible to approximately reconstruct low-body order features, while the opposite is not true. Even when using a non-linear (or locally-linear) mapping, reconstructing 3-body features from 2-body information is virtually impossible.
The 3-to-4-body mapping is more subtle: an overall reconstruction based on a linear model is not possible, but a local mapping works well, provided that the structures are far from the manifold of structures for which the 3-body description is not injective.
The associated transformation, however, is highly non-linear, and a kernel model that can reconstruct 4-body features shows poor transferability outside of the training set, which hints at similar shortcomings whenever one wanted to use it to learn a property that depends strongly on 4-body correlations.
Even though an overall linear reconstruction is not possible, the $\nu$-to-$(\nu+1)$-body mapping error decreases with increasing $\nu$, indicating that less information is added with higher body-orders.
This is consistent with the satisfactory results that have been obtained in the regression of atomistic properties using only 3-body information~\cite{glie+18prb, paru+18ncomm}, even though the fundamental incompleteness of 3-body features has been shown to have implications for the asymptotic learning performance~\cite{pozd+20prl}.
An analysis based on the GFRE might help determine the high-order correlations that provide the highest amount of information, and combined with an iterative scheme to evaluate the corresponding features~\cite{niga+20jcp} provide a strategy to increase model accuracy with an affordable computational effort.

We also investigate the effect of changing the metric used to compare features, by juxtaposing the Euclidean distance (that is induced by a linear description of the feature space) with a Wasserstein metric, that can be applied to the comparison of $n$-body correlation features when expressed as real-space distributions. We find that -- with an appropriate normalization -- the Wasserstein distance can be seen as a proxy of the minimal amount of distortion needed to transform an environment into another, and that this behavior induces smooth interpolation between sparse reference points, contrary to what is observed for the Euclidean distance.
However, both an aggressive smearing of the atom density, and the use of a more realistic data set cure the pathological behavior of the linear featurization, so that the Wasserstein metric should not be regarded as superior to the Euclidean one, but complementary. Generalizing the Wasserstein metric to higher body-order correlations, which induce a higher-dimensional feature space that is more likely to be sparsely populated, would be an interesting further research direction.

The analysis in this work can be extended to compare atom-density representation with a broader class of descriptors based on topological~\cite{isayev2017universal, sutton2019crowd, liu2019n} or physicochemical information~\cite{pilania2013accelerating, ward2016general,ouya+18prm} as well as property-dependent representations induced by neural network frameworks~\cite{schu+18jcp, cohen2018spherical}.
Even more broadly, an objective measure of the relative effectiveness of features can guide the development of machine learning schemes for any problem that depends strongly on the strategy used to obtain a mathematical description of its inputs.
The feature space reconstruction error and distortion can be incorporated into any machine learning frameworks to drive feature selection algorithms~\cite{imba+18jcp,ouya+18prm,Paleico2020arxiv} or to ensure that implementation choices that improve computational efficiency do not cause a degradation in the resolving power of the resulting features.
\section*{Acknowledgments}

AG and MC acknowledge support from the Swiss National Science Foundation (Project No. 200021-182057).
GF acknowledges support by the SCCER Efficiency of Industrial Processes, and by the European Center of Excellence MaX, Materials at the Exascale - GA No. 676598.
GI acknowledges support by the NCCR MARVEL,
funded by the Swiss National Science Foundation (SNSF).

\newcommand{\noopsort}[1]{}

\end{document}